\documentclass[useAMS,usenatbib]{mn2e}            
\usepackage{journals}
\usepackage[pdftex]{graphicx,color}       
\usepackage[latin1]{inputenc}
\usepackage{graphics}    
\usepackage{amsfonts}    
\usepackage{amsmath}
\usepackage{multicol}     
\usepackage{layout}     
\usepackage{amssymb}
\usepackage[a4paper,colorlinks=true,pdfstartview=FitV,
linkcolor=red,citecolor=blue,urlcolor=magenta]{hyperref}
\title[Cosmology through arc statistics I] {Cosmology through arc
  statistics I: sensitivity to  $\Omega_m$ and $\sigma_8$} 
\author[Boldrin et al.] {\parbox{\textwidth}{Michele Boldrin$^{1}$,
    Carlo Giocoli$^{2,3,4,5}$, Massimo Meneghetti$^{4,6}$, Lauro
    Moscardini$^{2,3,4}$, Giuseppe Tormen$^{1}$, Andrea Biviano $^{7}$} \\ \\  
$^{1}$  Dipartimento di Fisica e Astronomia, Universit\`a di Padova, vicolo  dell'Osservatorio  3,  35122 Padova,  Italy   \\ 
$^{2}$  Dipartimento di Fisica e Astronomia, Alma Mater Studiorum Universit\`a di Bologna, viale Berti Pichat 6/2,  40127  Bologna,  Italy  \\ 
$^{3}$ INFN - Sezione di Bologna, viale Berti Pichat 6/2, 40127 Bologna, Italy \\ 
$^{4}$ INAF - Osservatorio Astronomico di Bologna,  via Ranzani 1, 40127 Bologna, Italy\\
$^{5}$ Aix Marseille Universit\'e, CNRS, LAM (Laboratoire d'Astrophysique de Marseille), UMR 7326, 13388 Marseille, France \\
$^{6}$ Jet Propulsion Laboratory, 4800 Oak Grove Dr., Pasadena, CA
91109, USA\\
$^{7}$ INAF - Osservatorio Astronomico di Trieste, via G.B. Tiepolo 11, 34143 Trieste}

\begin{document}

\def\solmass{$h^{-1}\mbox{M}_{\odot}\;$}

\date{}
\maketitle
\label{firstpage}
\pagerange{\pageref{firstpage}--\pageref{lastpage}} \pubyear{}
\begin{abstract} 

  The next generation  of large  sky photometric  surveys will finally be
  able to use arc statistics as a cosmological probe.  Here we present
  the first  of a series of  papers on this topic.   In particular, we
  study  how  arc  counts  are  sensitive  to  the  variation  of  two
  cosmological  parameters:  the  (total)  matter  density  parameter,
  $\Omega_m$, and the normalisation  of the primordial power spectrum,
  expressed in  terms of $\sigma_8$.  Both  these parameters influence
  the abundances  of collapsed structures and their  internal structure.  We compute the
  expected number  of gravitational arcs with  various length-to-width
  ratios in mock light cones, by varying these cosmological parameters
  in the ranges  $0.1\leq\Omega_m\leq0.5$ and $0.6\leq\sigma_8\leq 1$.
  We find that  the arc counts dependence on  $\Omega_m$ and $\sigma_8$
  is  similar, but  not identical,  to that  of the  halo counts.   We
  investigate how the precision of the constraints on the cosmological
  parameters based on  arc counts depends on the survey  area. We find
  that the  constraining power  of arc statistics  degrades critically
  only for surveys  covering an area smaller than $10\%$  of the whole
  sky. Finally, we  consider the case in which the  search for arcs is
  done  only in  frames  where galaxy  clusters  have been  previously
  identified.   Adopting the  selection function  for galaxy  clusters
  expected  to  be  detected  from photometric  data  in  future  wide
  surveys, we find  that less than $10\%$ of the  arcs will be missed,
  with  only a  small  degradation of  the corresponding  cosmological
  constraints.

\end{abstract}
\begin{keywords}
 Arc  statistics;  strong   gravitational  lensing;  galaxy  clusters;
 cosmology.
\end{keywords}

\section{Introduction}

The  importance  of  galaxy  clusters   in  cosmology  is  well  known
\citep[for a  review, see][and references  therein]{allen2011}.  Being
the most massive bound systems in  the universe, they trace the latest
stage of structure  formation. Their abundance and mass  as a function
of redshift are thus highly indicative of how the growth of the cosmic
density  fluctuations occurs  and can  thus be  used to  constrain the
matter  content,  the initial  power  spectrum  normalisation and  the
expansion          history           of          the          universe
\citep{eke1998,borgani2001,reiprich02,
  allen2003,schuecker2003,henry2004,vikhlinin2009,mantz2010,rozo2010,sehgal11,
  planck2013_xx,benson2013,mantz2014}.

Moreover, galaxy clusters are  essential cosmic laboratories where the
complex interaction between baryons and  dark matter can be studied in
detail,       in      particular       during      merger       events
\citep{clowe2006,merten2011}.

Being the most massive structures in the Universe, galaxy clusters are
also      the       most      powerful       gravitational      lenses
\citep{lensinglectures,weaklensing,kneib2011}. In particular, they are
responsible  for highly  non-linear  lensing effects  taking place  in
their  densest  regions, i.e.   in  their  cores.  In  this  so-called
``strong"  lensing  regime,  the  images of  background  galaxies  are
heavily distorted,  often leading  to the appearance  of gravitational
arcs  with  large  length-to-width ratios  \citep[see][and  references
therein]{kneib2011,meneghetti13}.

The efficiency of  galaxy clusters to produce arcs with  a given ratio
between their length $l$ and width $w$ is quantified by means of their
strong lensing  cross section $\sigma_{l/w}$.  This is defined  as the
area on the source  plane where the source has to  be located in order
to  form an  arc with  such a  $l/w$ ratio.   To have  a large  strong
lensing cross  section, the  projected mass  distribution of  the lens
must  be exceptionally  dense on  the plane  of the  sky, also  called
``lens plane". Sometimes, this can be the result of projection effects
(halos elongated  along the  line-of-sight, superposition  of multiple
structures  at different  redshifts,  etc),  but, generally  speaking,
gravitational arcs trace the highest peaks in the cosmological density
field.  \citet{B98} first pointed out that counting gravitational arcs
may  be  a competitive  method  to  constrain cosmological  parameters
\citep[see e.g.][]{fedeli06,meneghetti13}.

Unfortunately, strong  lensing events  such as gravitational  arcs are
rare. Given  the relatively low number  of arcs discovered so  far and
the high inhomogeneity of cluster  optical surveys, the application of
arc statistics in cosmology has been attempted mainly with the goal of
possibly  falsifying the  standard cosmological  model (the  so-called
concordance   $\Lambda$CDM)   rather    than   actually   constraining
cosmological  parameters.   For  about  $15$  years,  scientists  have
debated  on the  existence or  nonexistence of  a tension  between the
observed number of  arcs and the predictions derived  in the framework
of the  cosmological model  favoured by observational  data \citep[see
][and references therein]{meneghetti13}.   Undoubtedly, arc statistics
suffered so far from the lack of suitable large observational datasets
for a  reliable comparison to theoretical  predictions.  The situation
is likely to change radically in the near future, thanks to the advent
of large  optical surveys,  covering areas in  the range  from several
thousands of square degrees \citep{kids,des2005}, to (almost) the full sky
\citep{lsst,hsc,EUCRB,wfirst}.

\begin{figure*}
 \centering
 \includegraphics[scale=0.55]{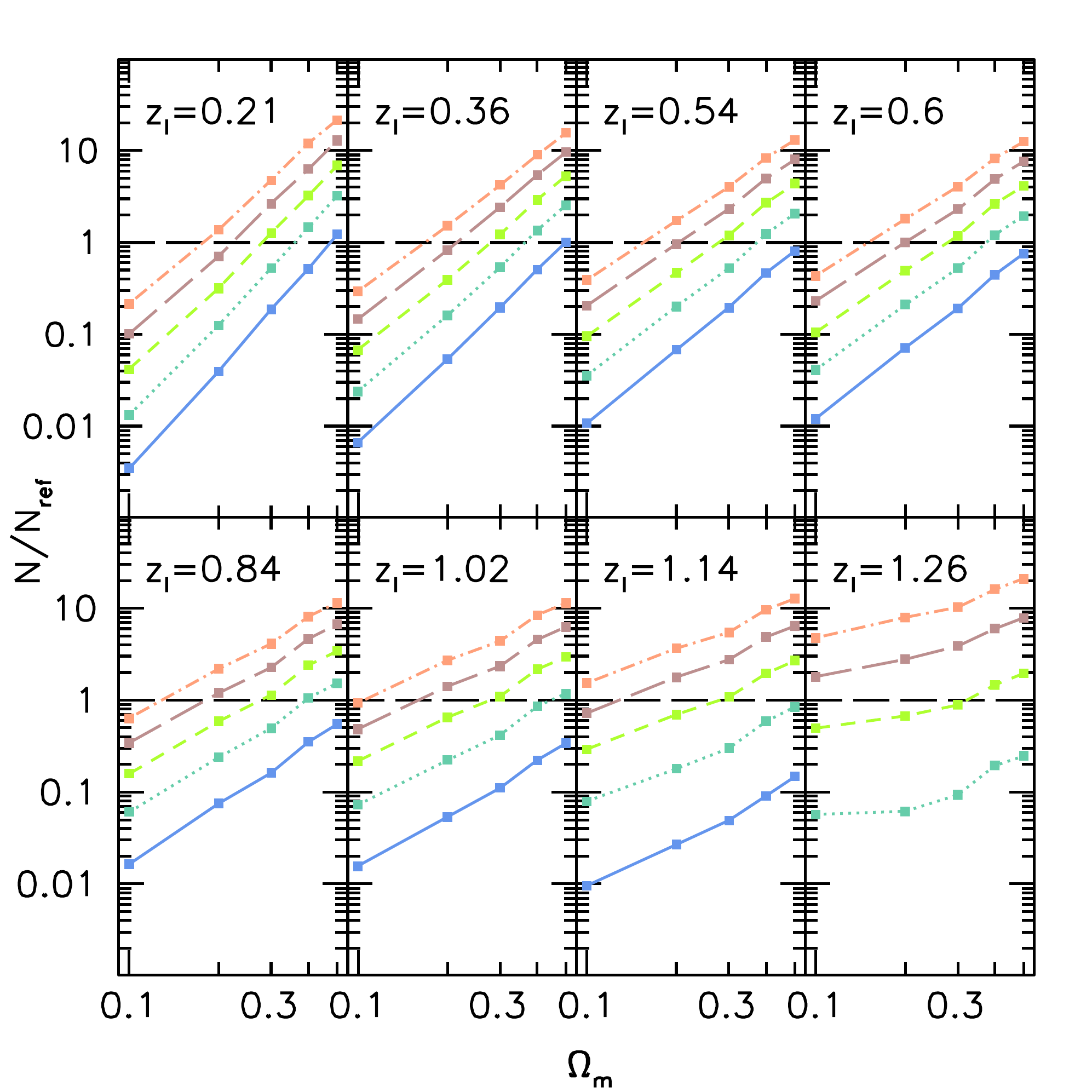}
 \caption{Number of arcs (normalised to the reference WMAP7 cosmology)
   as function of  $\Omega_m$ and for different  values of $\sigma_8$.
   Different panels refer to  different redshift bins between $z=0.21$
   and $z=1.26$, as labeled.  The reported counts represent the median
   of 128  different light-cone  realisations for each  combination of
   the cosmological parameters. Solid blue, dotted cyan, dashed green,
   long-dashed  brown and  dot-dashed dark  orange lines  indicate the
   results for $\sigma_8= 0.6$, 0.7, 0.8, 0.9 and 1, respectively. The
   results refer to arcs with  $l/w\ge 10$ and sources $1\sigma$ above
   the mean background noise level.}
 \label{fixedzl}
\end{figure*}

In \citet{boldrin2012} we forecasted  the number of gravitational arcs
visible in  the future wide  field survey to  be performed by  the ESA
Euclid mission \citep{EUCRB}.  A  further step is the analysis
of the sensitivity  of arc statistics on  cosmological parameters.  In
particular in this work, we focus  on the dependence of arc statistics
on  the  (total)  matter  density  parameter  $\Omega_m$  and  on  the
normalisation of the primordial power  spectrum, expressed in terms of
$\sigma_8$. More precisely, we investigate the region of the parameter
space  defined by  $\Omega_m=[0.1-0.5]$, and  $\sigma_8=[0.6-1.0]$. We
sample  the   parameters  at   intervals  $\Delta   \Omega_m=0.1$  and
$\Delta  \sigma_8=0.1$, thus  investigating  a total  of 25  different
cosmological models, always  making the assumption of  a flat Universe
($\Omega_m +  \Omega_{\Lambda}=1$).  To help a  direct comparison with
the  results  of  \citet{boldrin2012},  here  we  assume  a  reference
cosmology    defined    by     the    parameters    ($\Omega_m=0.272$,
$\sigma_8=0.809$) with  present Hubble parameter  $H_0=70.4$ km/s/Mpc,
in agreement  with the  WMAP7 results \citep{komatsu2011}.  However we
will  discuss also  the results  obtained with  the assumption  of the
parameters recently derived by \citet{planckParameters}.

The paper is organized as follows. In Section \ref{method} we briefly
describe the method adopted to compute the number of arcs and to build
the mock catalogs.  In Section \ref{results} we present our main
results, i.e. the total number of arcs and their redshift distribution
as a function of $\Omega_m$ and $\sigma_8$.  Section \ref{discussion}
is devoted to a discussion of the origin of the cosmological influence
on arc statistics; we also consider how the results can change when
including the effects of possible systematics. In Section \ref{testbed} 
we discuss the agreement between the predictions obtained with our methodology and the arc counts in real surveys, using the recent results of Xu et al. (2015)
based on the CLASH survey.
Finally, in Section \ref{conclusions} we summarise and draw our conclusions.

\begin{figure*}
  \includegraphics[scale=0.35]{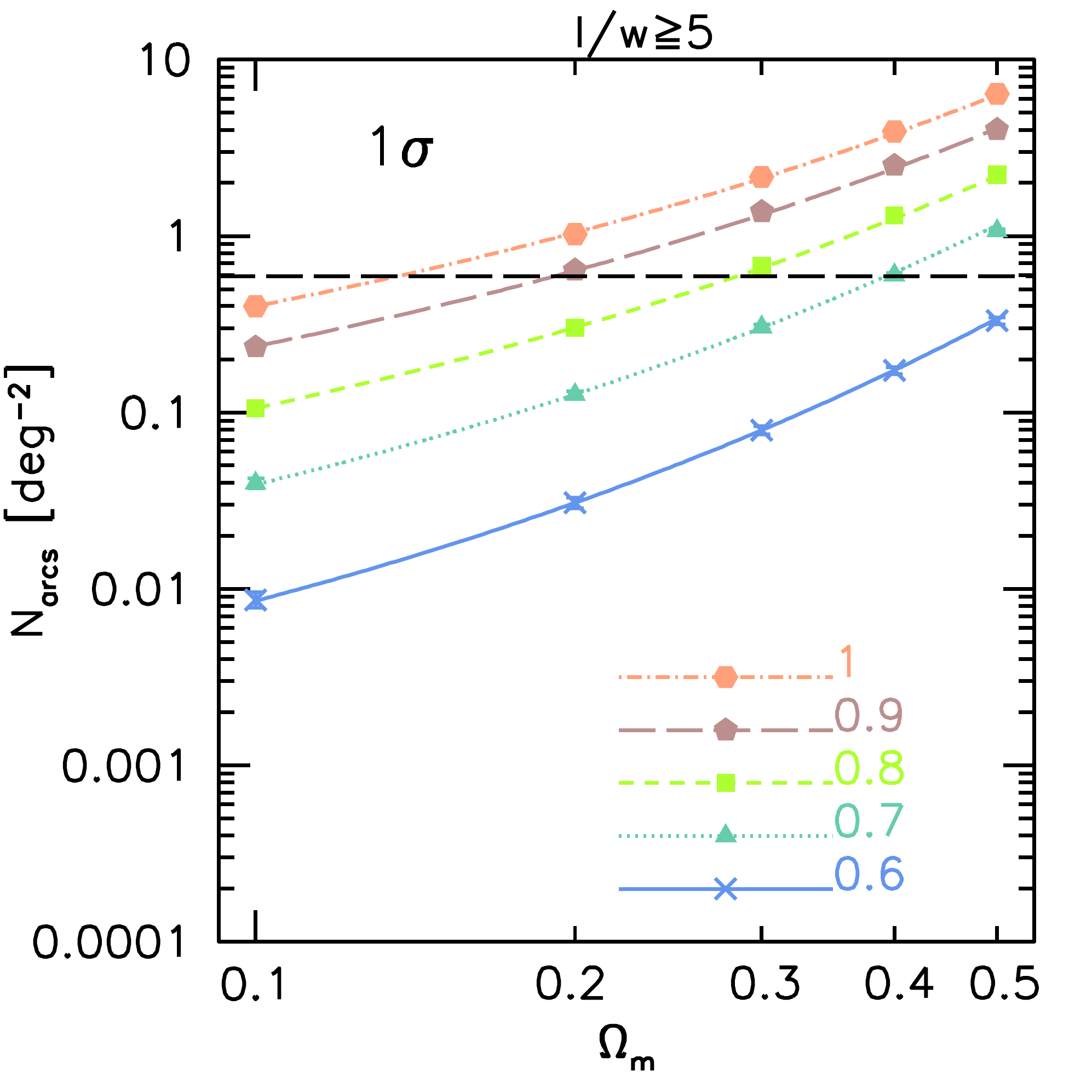}
 \includegraphics[scale=0.35]{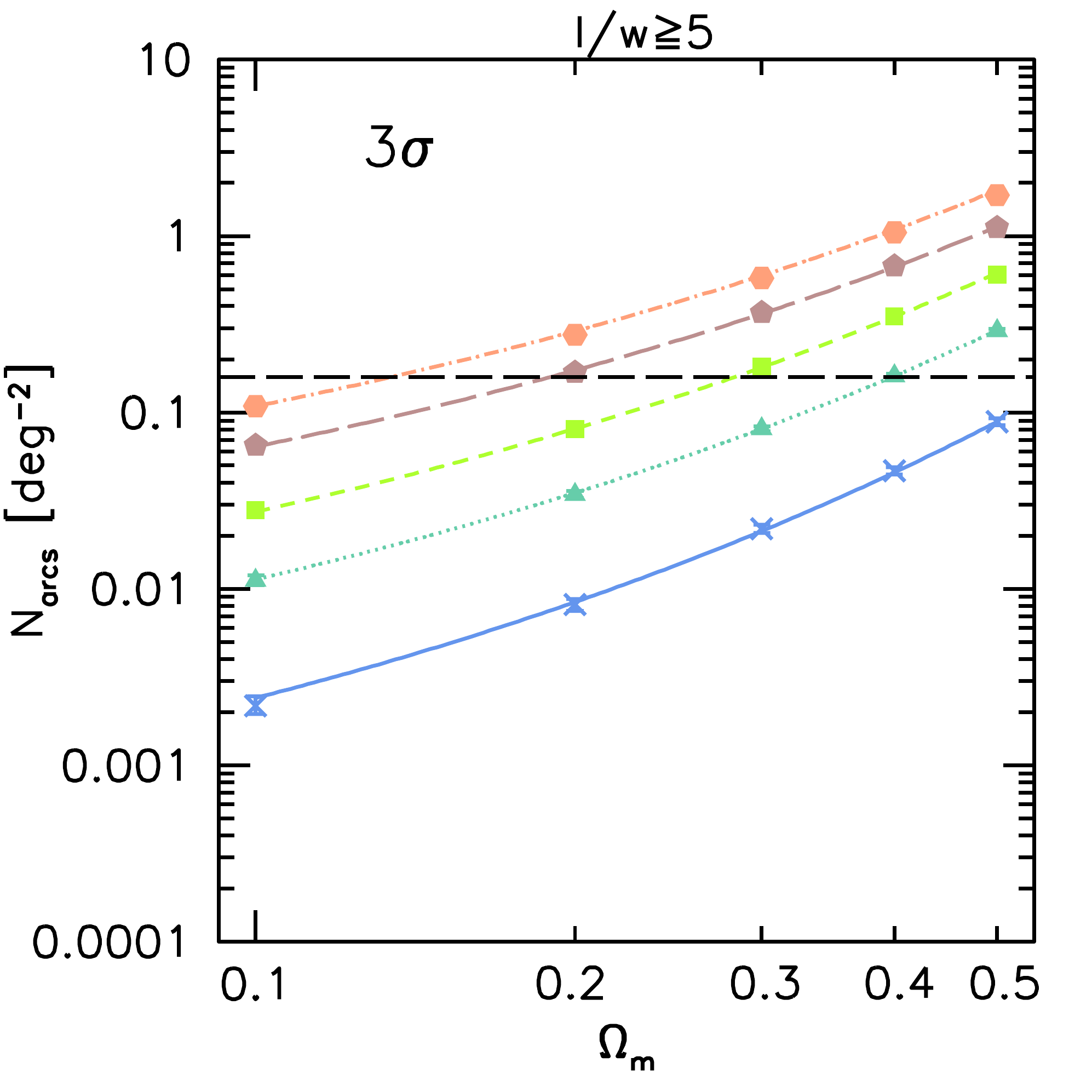}
 \includegraphics[scale=0.35]{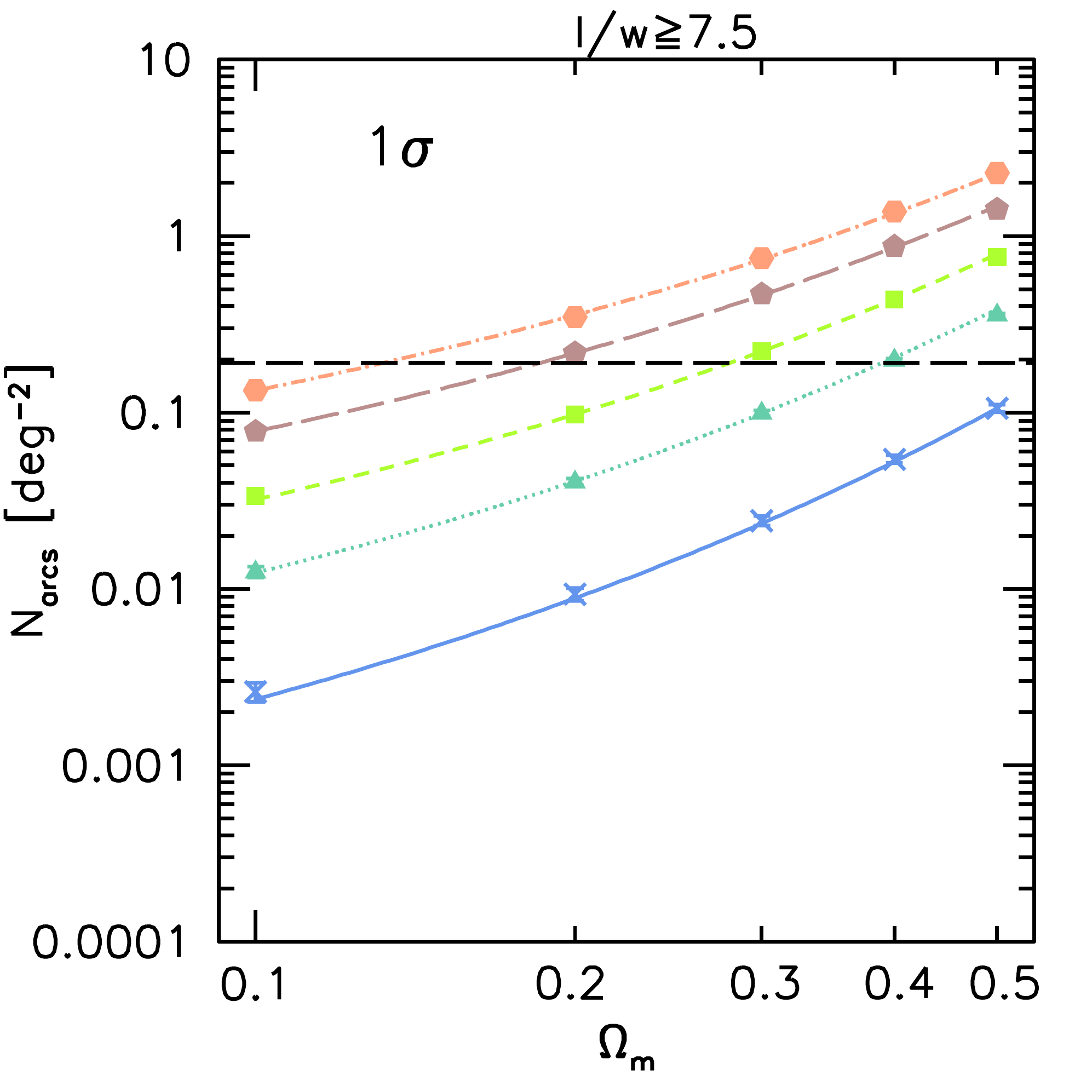}
 \includegraphics[scale=0.35]{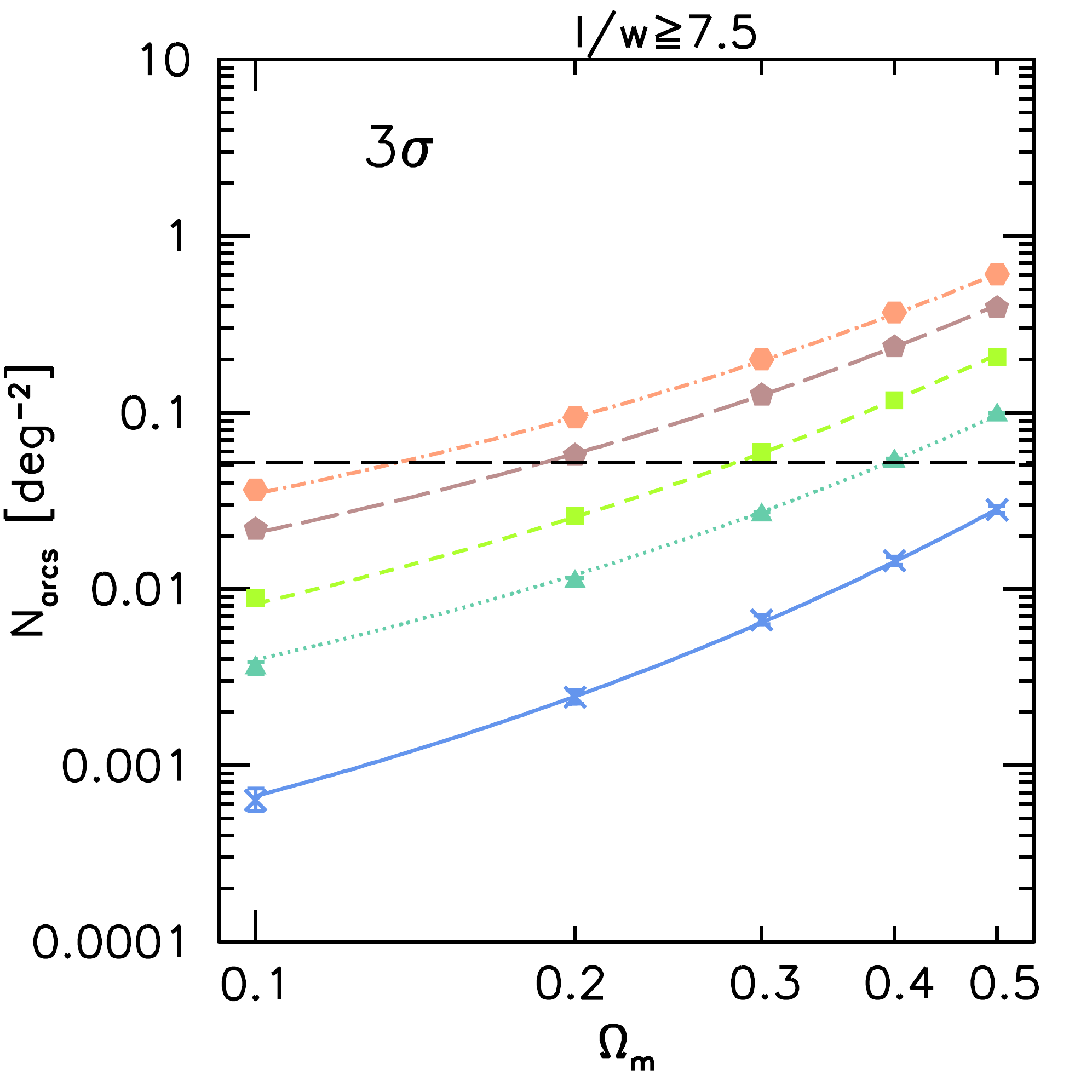}
 \includegraphics[scale=0.35]{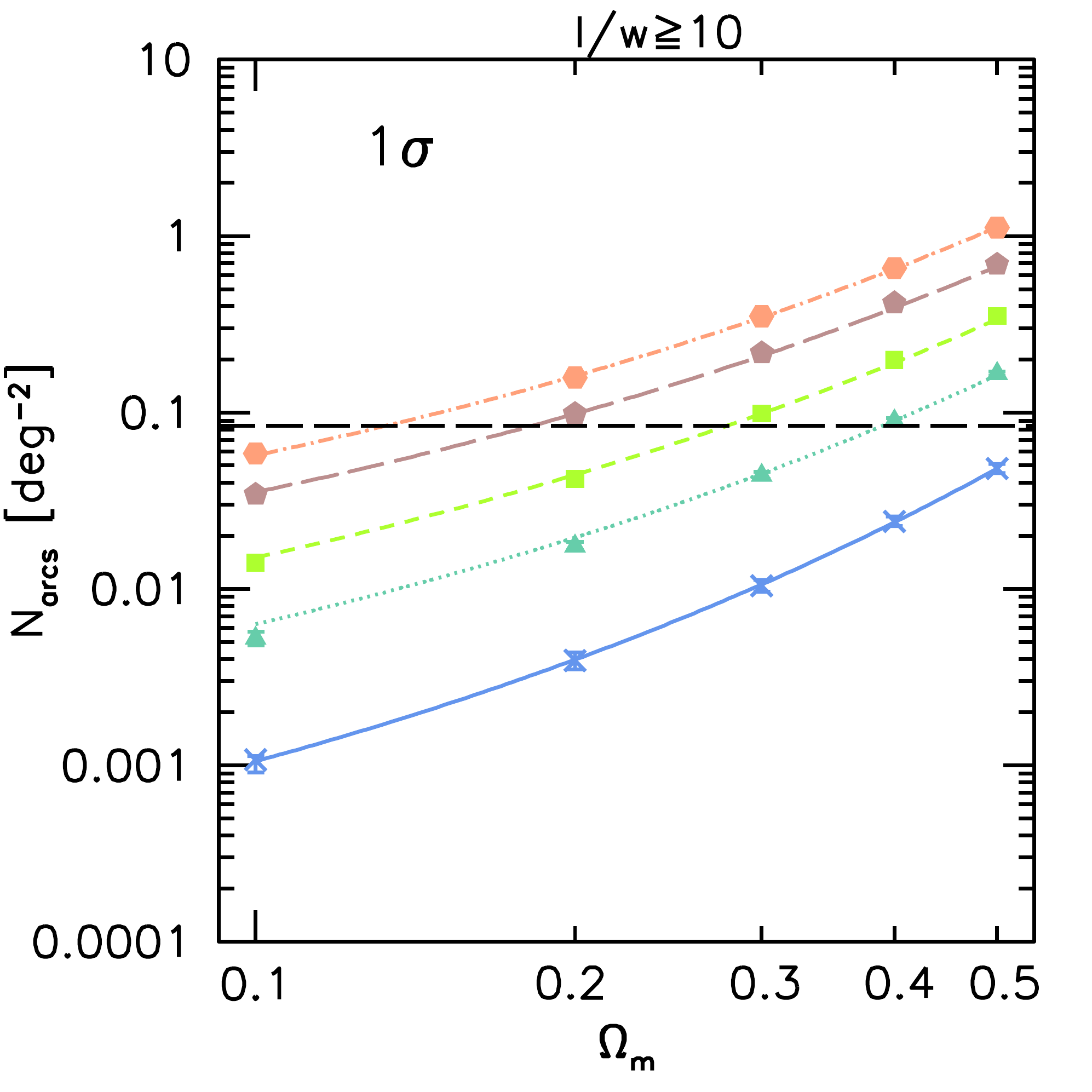}
 \includegraphics[scale=0.35]{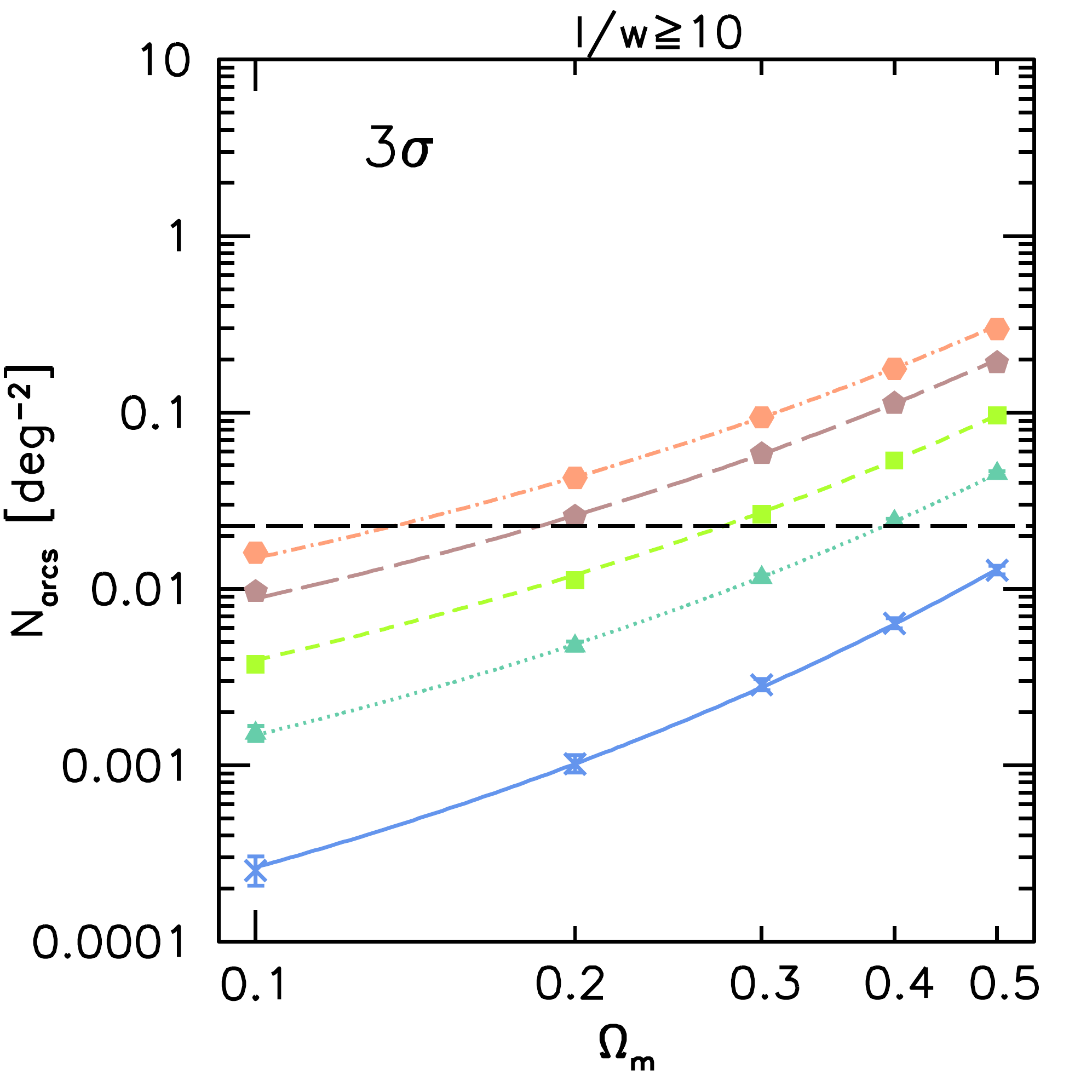}
 \caption{Number density of arcs as a function of $\Omega_m$ for
   different values of $\sigma_8$.  The right and left columns refer to
   sources detectable at $1\sigma$ and $3\sigma$ above the mean
   background noise level, respectively.  From top to bottom, the
   different panels show the results for three choices of minimum
   $l/w$, namely 5, 7.5 and 10.  Line and color styles are as in
   Fig.~\ref{fixedzl}. In each panel the horizontal dashed line shows
   the counts in the considered reference mode..}
 \label{fittot2}
\end{figure*}

\section{Method}
\label{method}
In this section, we summarise the steps undertaken for calculating the
number of arcs produced by a population of strong lenses.

\subsection{Generation of lenses with {\tt MOKA}}
In  the attempt  of  properly modelling  all  the relevant  structural
properties of  the lenses  in our  calculations of  $\sigma_{l/w}$, we
make    use     of    the    pseudo-analytic     code    \texttt{MOKA}
\citep{moka2011,giocoli12b}.   This  code   allows  to  generate  mock
cluster-size  gravitational lenses  in  any desired  cold dark  matter
(CDM) scenario,  including features  like: triaxiality  and projection
effects, scatter in concentration,  substructures, and the presence of
a  brightest central  galaxy, including  the effects  that its  growth
produces  adiabatic  contraction  on   the  cluster  dark-matter  halo
profile.   As  discussed  in  several earlier  works  based  on  fully
numerical                    simulations                    \citep[see
  e.g.][]{meneghetti2003CD,torri2004,meneghetti2007},     all    these
features play an important role for determining the cluster ability to
produce giant  arcs.  The  most important novelty  of our  work mainly
lies on the adoption of very realistic simulated strong lenses, as the
{\tt MOKA} code allows to do.

More  in detail,  to generate  the mock  lenses we  use the  following
prescriptions:
\begin{itemize}
\item the host halo mass density  profile is described by the Navarro,
  Frenk and White (NFW) radial function \citep{navarro1996};
\item the dark matter halo concentration is derived from the mass-concentration
  relation proposed by \citet{zhao2003}, assuming log-normal scatter
  of $\sigma_{\ln c}=0.25$;
\item the lens halo is triaxial and the axis ratios are generated
  assuming the triaxial distributions derived by \citet{jing2002};
  \item the spatial orientation is randomly chosen;
  \item the substructure abundance follows the sub-halo mass function
    derived by \citet{giocoli2008};
  \item the substructure radial  distribution follows that proposed by
    \citet{gao2004};
  \item the  substructures are  modelled as Single  Isothermal Spheres
    (SISs) \citep{metcalf01};
  \item the mass density profile of the Brightest Cluster Galaxy (BCG)
    resembles the Hernquist's profile \citep{hernquist1990}.
\end{itemize}
The main outcome of the \texttt{MOKA} code is the deflection angle map
of  each mock  halo.  From  the latter,  we obtain  $\sigma_{l/w}$ for
sources at $z_s=2$ via ray-tracing technique.  For simplicity,
  we  consider only  isolated lenses,  avoiding the  cases of  merging
  haloes,  even  though some  studies  revealed  the importance  these
  events  have  on   the  amplitude  of  the   strong  lensing  signal
  \citep[see e.g][]{torri2004}.   Furthermore, another  source of  enhancement of
  the signal is  given by the large scale structure  present along the
  line of  sight, which boosts the  lens projected mass on  the source
  plane.   Since beyond  purpose of  this paper  we postpone  detailed
  studies of those two aspects to a future work.

\begin{figure*}
 \centering
 \includegraphics[scale=0.75]{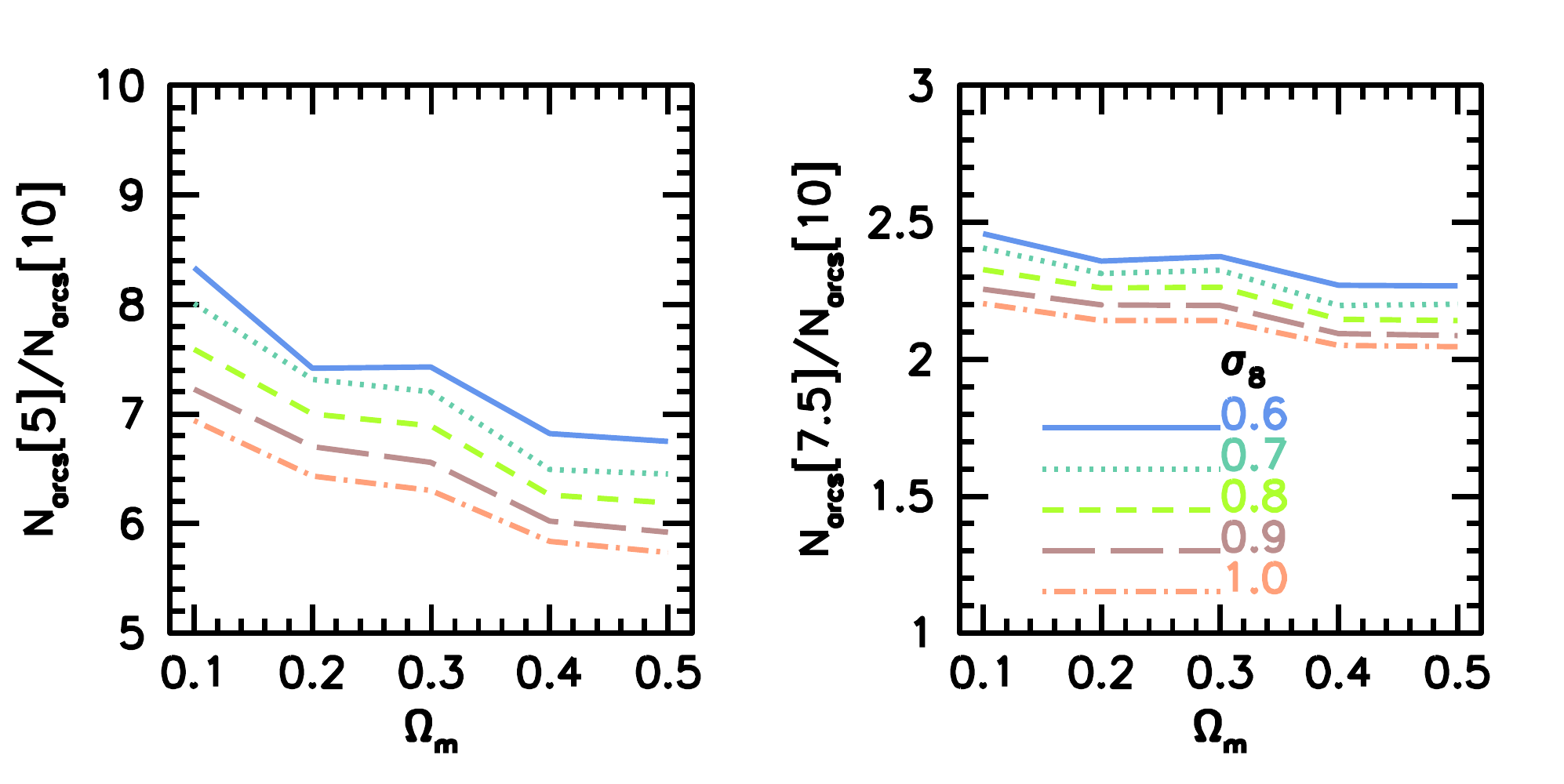}
 \label{ratios}
 \caption{The  abundances  of arcs  with  $l/w\ge 5$  (left panel)  and
   $l/w\ge 7.5$ (right panel)  relative to the abundances  of arcs with
   $l/w\ge 10$ as  a function  of $\Omega_m$.  Line  and color styles are as in Fig.~\ref{fixedzl}.
   The results refer to arcs detectable at $1\sigma$   above  the  mean  background  level. 
\label{figratios}}
\end{figure*}

\subsection{Number of giant arcs}

We trace light-rays  through a uniform grid on the  lens plane towards
the source  plane, accounting for the  position-dependent deflections.
The source plane is populated with elliptical sources distributed on a
regular grid.   Their lensed  shapes are  recovered by  collecting the
light  rays hitting  them.  The  lengths and  widths of  the resulting
gravitational arcs are measured using  the method described in several
earlier  papers  \citep[see  e.g.][]{meneghetti2003CD}. As in \citet{boldrin2012}, the source population has random ellipticity and an apparent size depending on the redshift (see \citealt{boldrin2012}, Fig. 3.). To  properly
sample the  region of the  source plane,  where sources are  lensed as
giant arcs, we iteratively increase the resolution of source grid near
the lens  caustics. Each source is  then representative of an  area on
the  source  plane defined  by  the  local  resolution of  the  source
grid. Using this area as weight, we compute $\sigma_{l/w}$, defined as
the area on the  source plane within which a source  has to be located
for   being   lensed   as   a   giant   arc   with   a   given   $l/w$
\citep{meneghetti2008}.

The  number of  arcs  with length-to-width  ratio  larger than  $l/w$,
produced by  a lens  of mass  $M$ at redshift  $z_l$, can  be computed
solving the following relation:
\begin{equation}
  N_{l/w}(M,z_l)=\sigma_{l/w}(2)\int_{z_l}^{z_{s,max}}\mbox{d}z_sf_{\sigma}(z_l,z_s)n(z_s,S)\,,
  \label{N1halo}
\end{equation}
where $\sigma_{l/w}(2)\equiv \sigma_{l/w}(M,z_l,z_s=2)$  is the strong
lensing cross section for giant  arcs for sources at redshift $z_s=2$,
and  $f_{\sigma}\equiv \dfrac{\sigma_{l/w}(z_s)}{\sigma_{l/w}(z_s=2)}$
accounts  for the  scaling of  $\sigma_{l/w}$ with  $z_s$.  In
  order to  minimize the  computational time,  we recover  the scaling
  function  $f_{\sigma}$  computing   $\sigma_{l/w}(z_s)$  of  a  lens
  subsample,  adopting  different  values  of  $z_s$  \citep[see  for  more
    details][]{boldrin2012}.

A  crucial ingredient  is  the redshift  distribution  of the  sources
exceeding a given surface brightness  $S$, $n(z_s,S)$. We derive it by
simulating    an    observation    of     the    galaxies    in    the
Hubble-Ultra-Deep-Field   (HUDF)  using   the  \texttt{SkyLens}   code
\citep{meneghetti2008}, to  a depth which  is reasonable for  a future
wide field survey  from space. As a reference, we  consider the Euclid
wide survey, which is expected  to reach an average limiting magnitude
$m_{riz}=24.5$  \citep{EUCRB}.   Adopting typical  Euclid-like
exposure  times  and  background   levels,  we  estimate  the  surface
brightness limits  corresponding to $1\sigma$ and  $3\sigma$ above the
mean background level, and use  the simulated observations of the HUDF
to measure $n(z_s,S)$. Using a subset of {\tt MOKA} lenses to simulate
observations of gravitational arcs, we find that we can safely neglect
from the giant arc counts sources above $z_{s,max}=6$.

For each combination of cosmological  parameters, we produce a catalog
of cluster-sized lenses with different masses and redshifts. We define
100  mass  bins  which  are  uniformly  spaced  in  logarithm  between
$10^{13}$  and $10^{16}\;h^{-1}M_\odot$  and 8  redshift bins,  having
$\Delta z=0.03$ and centered at redshifts 0.21, 0.36, 0.54, 0.6, 0.84,
1.02, 1.14,  and 1.26. The choice  of such redshift bins  is optimised
for the expected  redshift distribution of the  lenses producing giant
arcs,  which  we  derived  in \citet{boldrin2012}  for  a  Euclid-like
survey. For each combination of redshift and mass we use {\tt MOKA} to
generate 100  halos with  different structural properties  and measure
their $\sigma_{l/w}$,  from which we  can derive the number  of giant
arcs they produce, as discussed above.

The catalog  of lenses is  then used  to generate 128  realisations of
lens distributions  (light-cones). In each light-cone,  which subtends
an area of 15,000 square degrees, we calculate the number of lenses of
mass  $M$ and  redshift $z_l$  according to  the \citet{sheth99}  mass
function, and  estimate the  total number  of arcs  by summing  up the
contributions  from each  individual  lens.  Finally,  we combine  the
different light-cones to measure the  median number of arcs per square
degree  and the  relative  scatter  as a  function  of the  considered
cosmological parameters.

\section{Results}
\label{results}

\subsection{Number of arcs as a function of redshift}
\label{numredshifts}

We begin  by discussing  how the  arc counts change  as a  function of
cosmology in  different redshift  bins. In Fig.~\ref{fixedzl}  we show
the number  of arcs, normalized  to the reference WMAP7  cosmology, as
function  of  $\Omega_m$. The  different  panels  refer to  the  eight
redshifts where the calculations  were performed. Different colors and
line styles  are used  to display  the results  for several  values of
$\sigma_8$: solid  blue, dotted cyan, dashed  green, long-dashed brown
and dot-dashed  dark orange lines  refer to $\sigma_8$=0.6,  0.7, 0.8,
0.9  and   1,  respectively.    Long-dashed  black   horizontal  lines
correspond to unity,  i.e. to the reference cosmology.  The  lack of a
blue  solid line  in the  last  panel is  due to  the inefficiency  of
clusters at $z_{l}=1.26$  to produce giant arcs in  the cosmology with
$\sigma_8=0.6$. As expected, at all  redshifts, the arc counts grow as
a function of  $\Omega_m$ and as a function  of $\sigma_8$, indicating
that  the  abundance   of  giant  gravitational  arcs   is  higher  in
cosmological models with  more matter and higher  normalisation of the
power spectrum of the primordial density fluctuations.

We also notice that the change of arc counts as a function of
cosmology depends on the lens redshift. The dependence on $\Omega_m$
is stronger at lower redshift, and flattens off as $z_l$ increases. On
the contrary, it appears that the value of $\sigma_8$ affects the
results more significantly at high redshift.

While the results in Fig.~\ref{fixedzl} refer to arcs with $l/w\ge10$
and sources above the $1\sigma$ background level, the trends remain
similar for other $l/w$ ratios and detection limits.

\subsection{The total number of arcs}
\label{totnum}
From  the distributions  obtained  from the  128 different  light-cone
realizations, we measure  the median number of arcs  per square degree
expected in each cosmological model.  This has been done by performing
a spline interpolation through the above-mentioned 8 redshifts up to a
maximum lens redshift of $z_l=1.5$.

In the  reference WMAP7  cosmology, the  expected number  densities of
arcs   per  square   degree  with   $l/w\ge  5$,   7.5,  and   10  are
$0.594\pm0.016$,  $0.194\pm0.006$, and  $0.085\pm0.003$, respectively.
These  are  in excellent  agreement  with  our estimates  reported  in
\citet{boldrin2012},  although  these  were obtained  using  a  larger
number of redshift bins and avoiding the interpolation.

In Fig.~\ref{fittot2}, we show the median arc number counts per square
degree as a function of $\Omega_m$. We also show how the counts vary
by changing the value of $\sigma_8$, using the same color and line
styles used in Fig.~\ref{fixedzl}.  From top to bottom, we show the
results for $l/w\ge 5$, 7.5, and 10, respectively. The left and the
right panels refer to detections at $1$- and $3\sigma$ above the level
of the background. Obviously, the results show the same dependence on
$\Omega_m$ and $\sigma_8$ reported in Fig.~\ref{fixedzl}.

We also  see that  the ratios  between counts  of arcs  with different
$l/w$   depend  on   the   cosmological  parameters.    As  shown   in
Fig.~\ref{figratios}, for  low $\Omega_m$, the abundance  of arcs with
$l/w\ge 5$ or $l/w\ge 7.5$, relative to that of arcs with $l/w\ge 10$,
is higher,  indicating that  halos in  these cosmological  models have
smaller critical lines and are  thus less efficient at producing large
distortions.   The ratios  also depend  on $\sigma_8$;  in cosmologies
with higher $\sigma_8$ halos are able to produce a higher abundance of
arcs  with large  $l/w$.  The  results  (here shown  only for  sources
$1\sigma$  above the  mean background  level) are  insensitive to  the
assumed detection  limit.  Therefore,  in the following  discussion we
will  show the  results  only  for arcs  detectable  at the  $1\sigma$
level. We will also focus on arcs with $l/w\ge 10$.

\begin{figure}
 \centering
 \includegraphics[scale=0.465,angle=270]{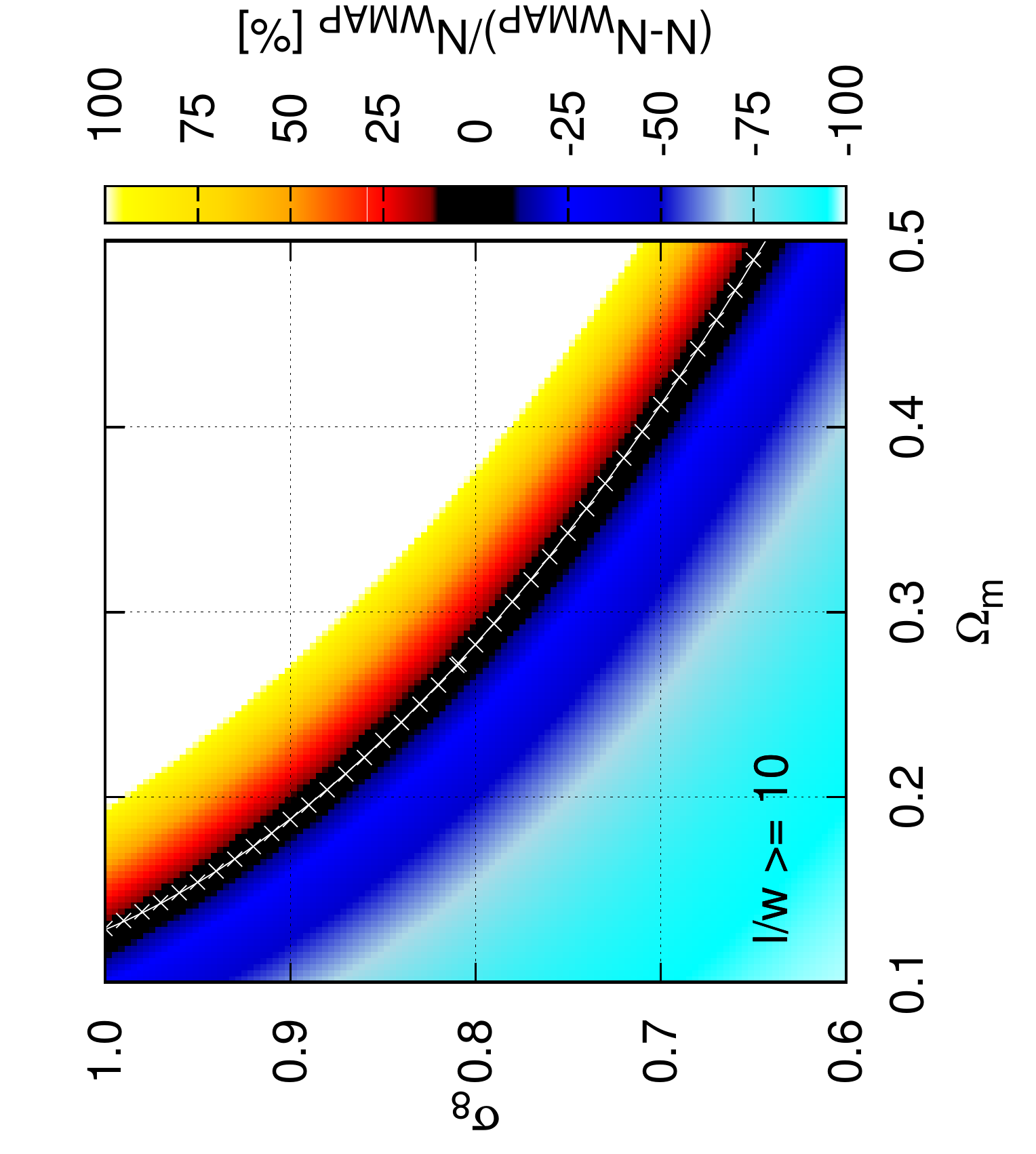}
 \includegraphics[scale=0.4]{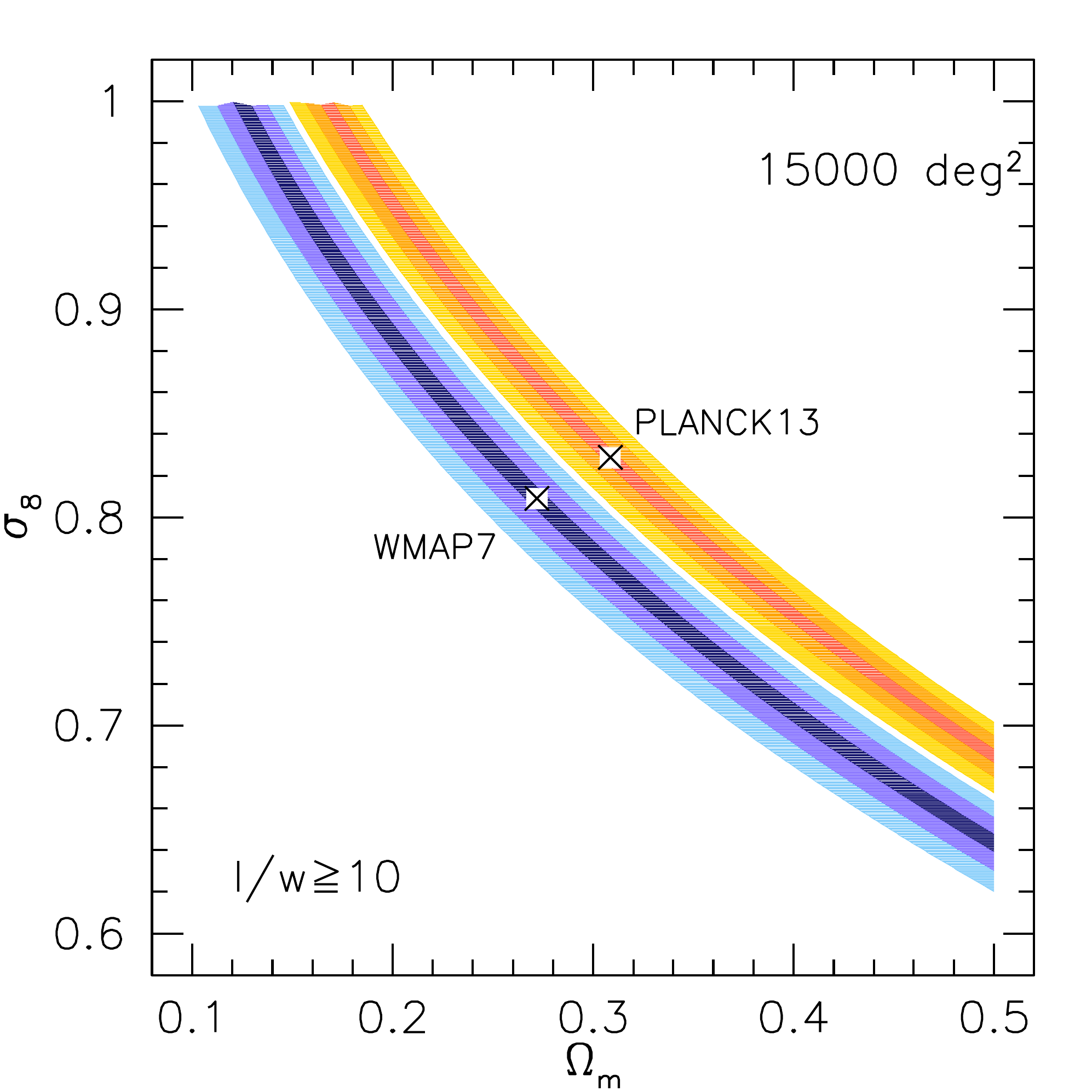}
 \caption{\textit{Upper  panel:} difference  in  the  arc counts  with
   respect to the reference WMAP7 cosmology in the $\Omega_m-\sigma_8$
   plane. The results  are shown for arcs with  $l/w\ge 10$ detectable
   at  $1\sigma$  above  the  background  level.   The  white  crosses
   represent the cosmological models having the same arc counts as the
   reference  WMAP7  model [relation~(\ref{totfit})].   \textit{Bottom
     panel:} levels  corresponding to  1, 3, and  5$\sigma$ deviations
   (from dark  to light colors) from  the WMAP7 (blue) and  the Planck
   (yellow) cosmologies  in the $\Omega_m-\sigma_8$ plane,  assuming a
   15,000  deg$^2$ survey  to the  expected depth  of the  Euclid wide
   survey.  The  crosses indicate  the position  of the  two reference
   models.}
 \label{constraints}
\end{figure}

The upper panel in Fig.~\ref{constraints}  shows the difference in the
arc  counts  relative   to  the  reference  WMAP7   cosmology  in  the
$\Omega_m-\sigma_8$ plane. Within the ranges explored in this work, we
may find  differences of up to one  order of magnitude for  the predicted
arc  counts  between cosmological  models.  We  also notice  that  the
cosmological parameters $\Omega_m$ and  $\sigma_8$ are degenerate with
respect to the arc counts. Indeed, the same number of arcs is expected
in cosmologies whose combination of  $\Omega_m$ and $\sigma_8$ lays in
a banana-like region extending from the upper left to the bottom right
corner of  the plane.  The  origin of  this degeneracy will  be better
discussed  in  Section~\ref{CosmInfl}.  Interestingly,  a  Planck-like
cosmology     with     $\Omega_m=0.3086$     and     $\sigma_8=0.8288$
\citep{planckParameters} produces $54\%$ more  arcs than the reference
WMAP7 cosmology.
  
\begin{figure}
 \centering
 \includegraphics[scale=0.4]{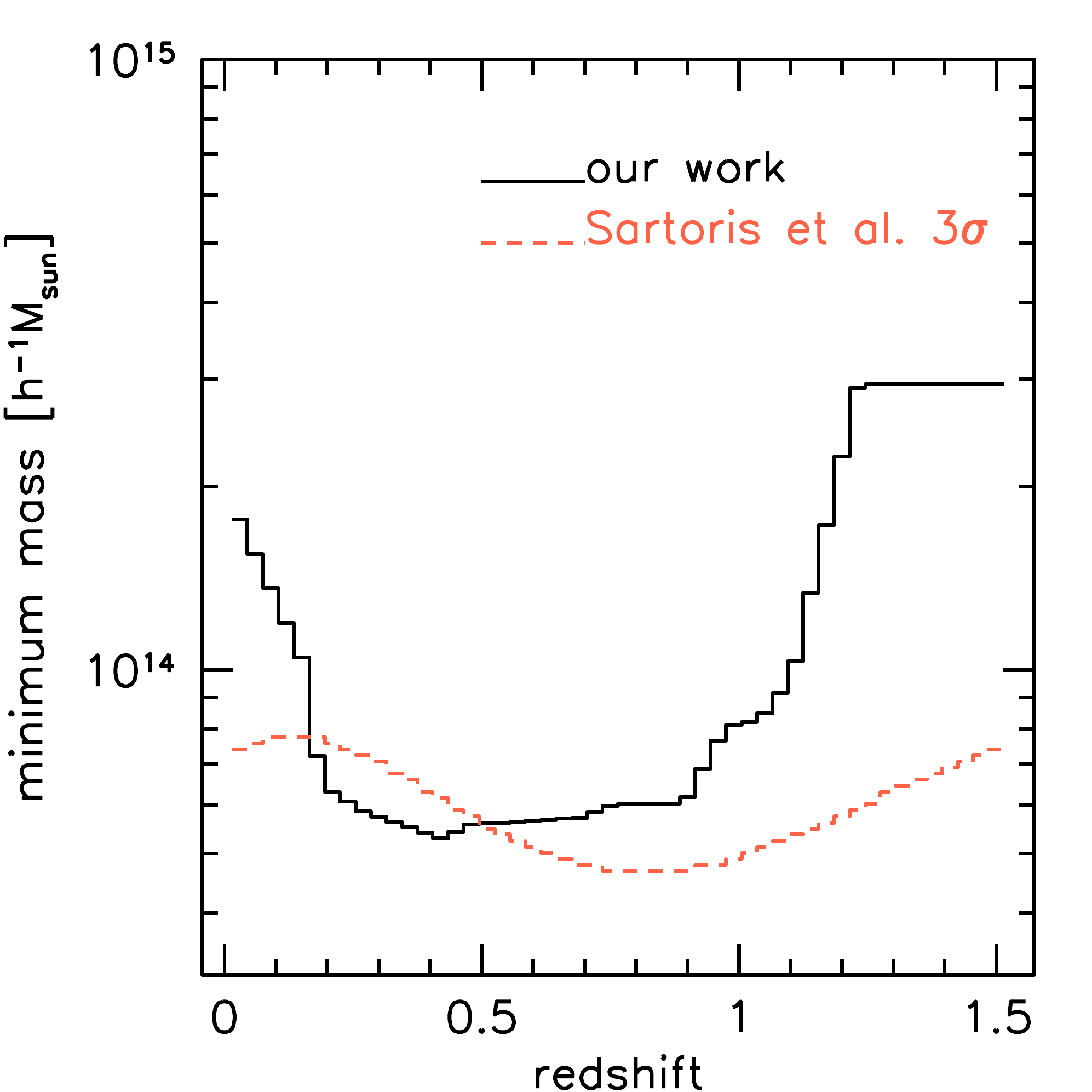}
 \caption{The strong  lensing selection function (black  solid curve),
   i.e. the minimum  galaxy cluster mass expected  to produce critical
   lines       for       sources        located       at       $z_s=2$
   \citep{meneghetti10a,boldrin2012}.  For comparison,  the red dashed
   curve  represents the  minimum mass  of galaxy  clusters which  are
   expected to  be detected  above three  times the  rms of  the field
   galaxy counts in the Euclid photometric survey \citep{sartoris15}.}
 \label{contBIV0}
\end{figure}

We find that in the case of the reference WMAP7 model, 
the equation describing the degeneracy curve between the
cosmological parameters has the following form:
\begin{equation}
  \Omega_m=A\sigma_8^2+B\sigma_8+C,
 \label{totfit}
 \end{equation}
 where $A=1.771$, $B=-3.952$ and $C=2.31$. Such function is given by
 the white line in the upper panel of Fig. \ref{constraints}.

\begin{figure}
 \centering
 \includegraphics[scale = 0.45]{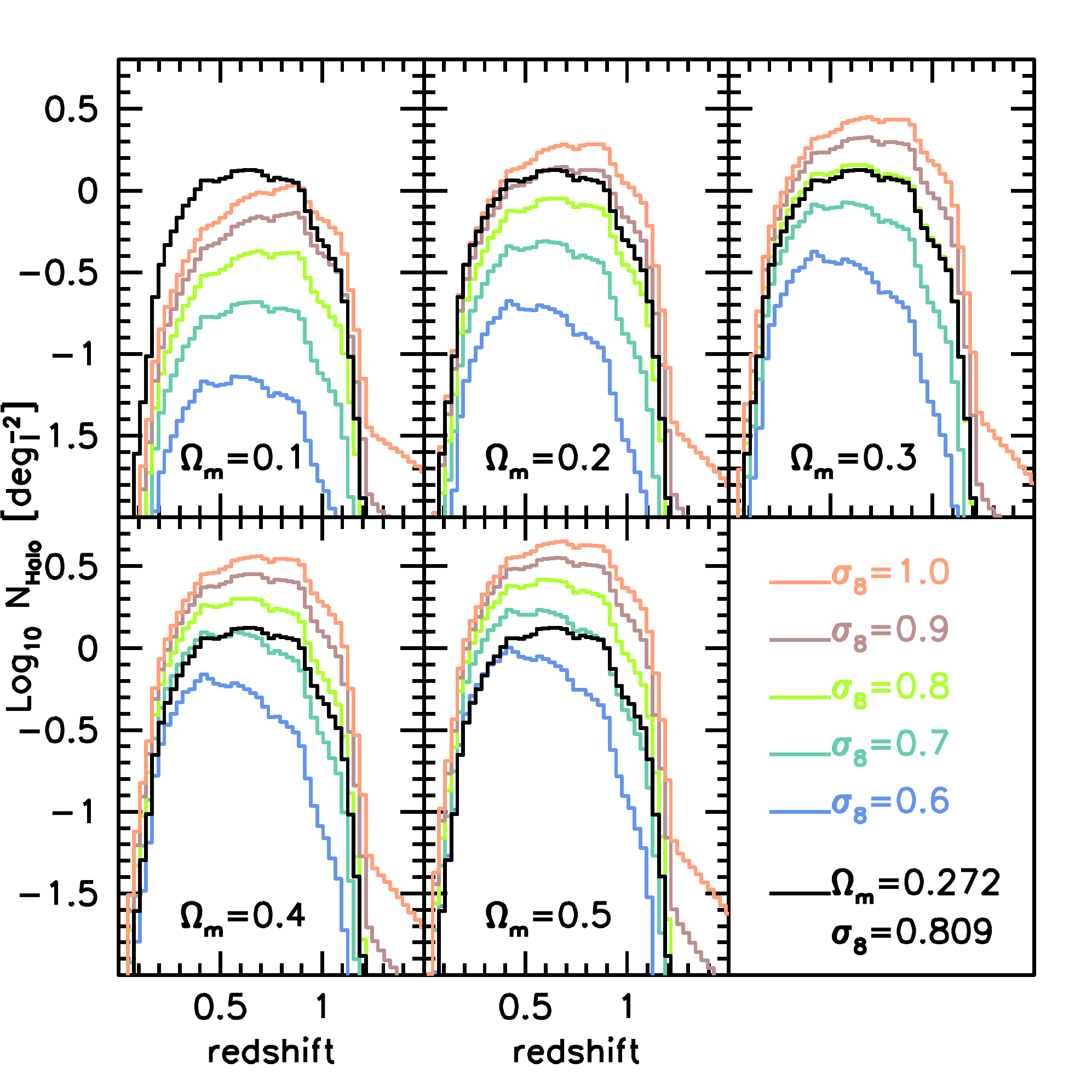}
 \caption{Number density  of expected strong  lenses as a  function of
   redshift, for cosmologies with different $\Omega_m$ and $\sigma_8$.
   Plots from left to right and  from up to bottom refer to increasing
   values  of  $\Omega_m$.   Different  colors  represent  counts  for
   various values of $\sigma_8$, as  labeled on the bottom right.  The
   black  line shown  in all  panels  represents the  results for  the
   reference WMAP7 model.}
 \label{halocounts}
\end{figure}

\begin{figure}
\hspace{-0.5cm}\includegraphics[scale     =      0.55,     angle     =
  270]{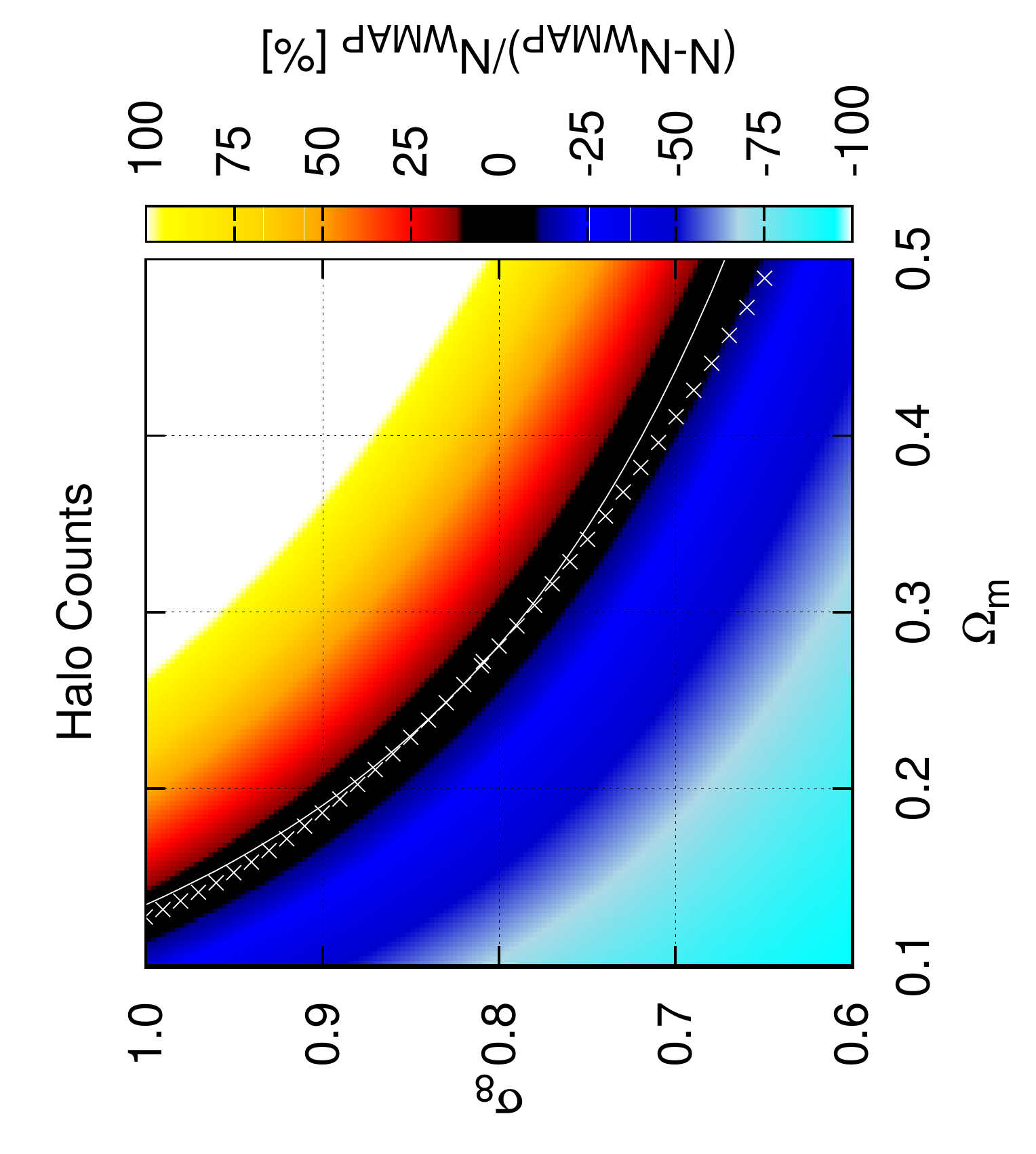}
  \caption{Relative difference of halo counts on the
    $\sigma_8$-$\Omega_m$ plane with respect to the reference WMAP7
    model.  The white crosses represent the degeneration curve
    relative to the arc counts, while the white solid line refers to the
    degeneracy curve for halo counts.}
\label{contourMF}
\end{figure}

In the attempt to quantify the uncertainty in the arc counts, we
define the 1$\sigma$ uncertainty on the number counts as
$\sigma\equiv(\sigma_{CV}^2+\sigma_P^2)^{1/2}$, where $\sigma_{CV}$ is
the cosmic variance, which is estimated from the 16th and 84th
percentiles of the distributions derived from the 128 light-cone
realisations of each tested cosmological model.  The other term
appearing in the equation, $\sigma_{P}\equiv\sqrt{N}$, is the
associated Poisson noise on the number counts.

In the bottom panel of Fig.~\ref{constraints}, we perform an error
analysis showing the levels corresponding to 1, 3, and 5 $\sigma$
deviations (from dark to light colors) from the WMAP7 and the Planck
cosmologies in the $\Omega_m-\sigma_8$ plane. The results were
obtained assuming a survey covering $15,000$ sq. degrees of the
sky to the depth expected for the Euclid mission.  It is interesting
to notice that a survey with the Euclid characteristics will be able
to distinguish these two cosmological models at more than $5\sigma$ level.

\section{Discussion}
\label{discussion}
\subsection{Influence of the cosmological parameters on arc statistics}
\label{CosmInfl}
In this  section we will discuss  in more detail some  aspects of the
influence of $\Omega_m$ and $\sigma_8$ on arc statistics.  In general,
the  cosmological  parameters  play  an  important  role  in  arc
statistics  through  the lens mass  function and  their  strong
lensing cross section, the latter depending on the geometry of the Universe and on
the structural properties of the lens halo.

In particular, the number of arcs is directly related to the number of
lenses able to produce arcs.  Following
\citet{meneghetti10a,meneghetti11}, this can be estimated including in
the mass function describing the lens distribution a sharp cut at the
minimum mass corresponding to the smallest systems in which we expect
to find critical lines for sources at $z_s=2$.  The
 shape of the adopted selection function as a
function of redshift is shown by the black curve in 
Fig.~ \ref{contBIV0} \citep[see also][]{boldrin2012}.

In Fig.~\ref{halocounts}, we present the number density (given per
square degree) of the lenses as a function of redshift. In each panel,
we keep fixed $\Omega_m$ as labeled and we vary the value for
$\sigma_8$, using the color code indicated on the bottom right.  To
facilitate the comparison, the lens number density in the reference WMAP7
cosmology is shown in black in all panels.  From the figure,
the strong effect of the different matter density on the
lens abundances and the anticipated structure formation originated by
a higher power spectrum normalization are clear.

In Fig.~\ref{contourMF}, adopting the same color code as in the upper
panel of Fig.~\ref{constraints}, we show the difference in the lens
counts relative to the reference WMAP7 cosmology in the
$\Omega_m-\sigma_8$ plane.  The white solid curve in the figure
represents the degeneracy between $\Omega_m$ and $\sigma_8$ for the
halo counts, for which we find the following relation:
\begin{equation}
 \sigma_8(\Omega_m/0.272)^{0.304} = 0.809\,.
 \label{halofit}
\end{equation}
Even if  with some differences,  this curve  is close to  the relation
(shown by the  white crosses) representing the degeneracy  we found in
the   $\Omega_m-\sigma_8$  plane   for  the   arc  counts   (see  also
Fig.~\ref{constraints}): this is clearly due to the fact that the most
important  ingredient  for arc  statistics  is  the lens  mass
function.   However,  if  one  compares the  amplitude  of  the  count
variation  by looking  at  the width  of the  coloured  strips, it  is
evident that  the arc  density is more  sensitive to  the cosmological
parameters than the  simple halo density: consequently,  a wide survey
of gravitational arcs could potentially give significant constraints.

\begin{figure}
 \centering
 \includegraphics[scale = 0.42]{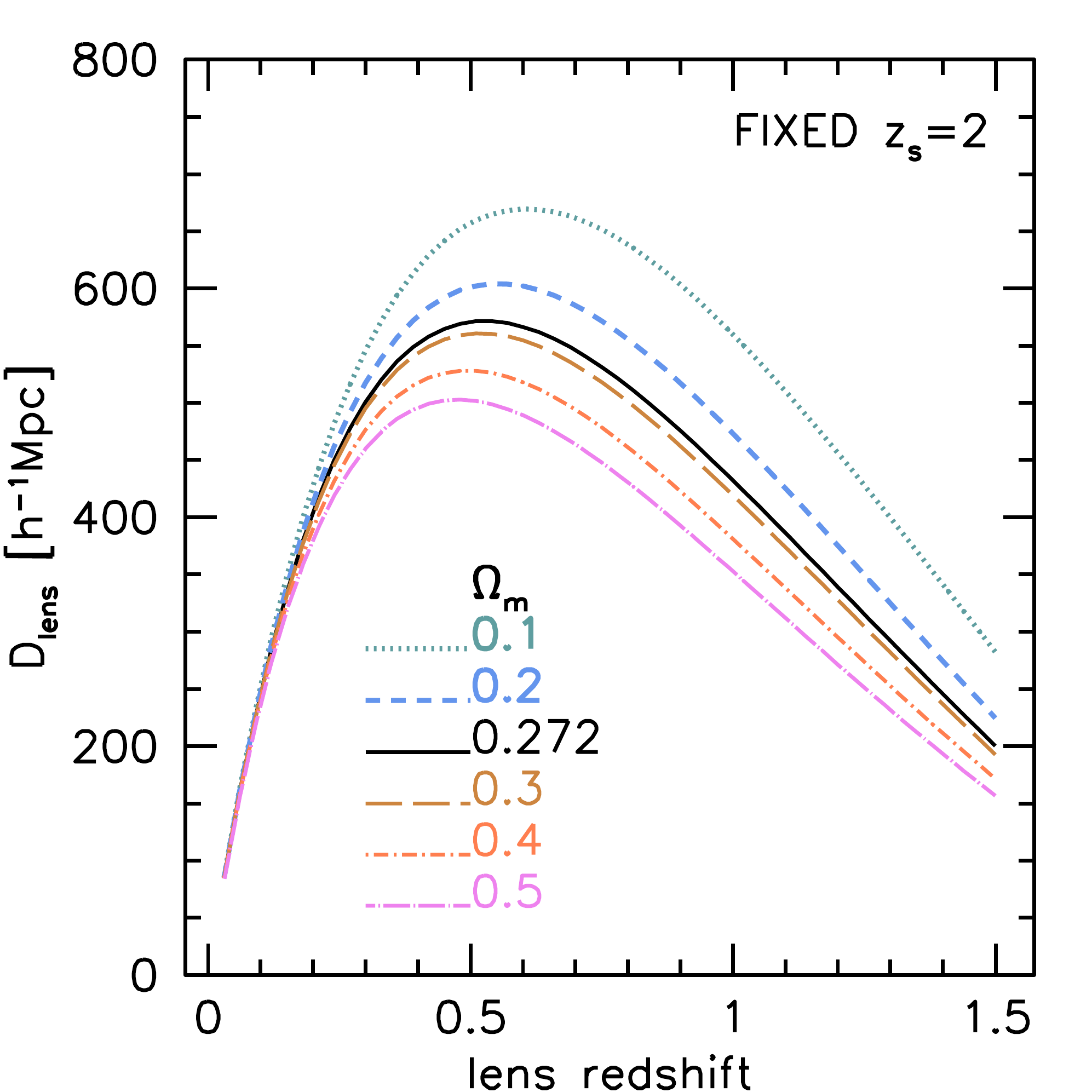}
 \caption{Lensing  distance for different values  of $\Omega_m$.
   Sources are kept fixed at redshift $z_s=2$.}
 \label{lensDist}
\end{figure}

The larger  sensitivity of arc  statistics is due to  the cosmological
dependence of the other main ingredients, such as the angular diameter
distances of lenses and sources and the lens structural properties. We
know that  the first condition  for an  axially symmetric lens  to act
like a strong lens  is that in some points $\vec x$  on the lens plane
the condition \begin{equation} \kappa(\vec x)>1
  \label{kappa}
\end{equation}
occurs, where $\kappa\equiv \Sigma(\vec x)/\Sigma_{cr}$ is the
so-called convergence, $\Sigma(\vec x)$ is the lens projected mass
density and
\begin{equation}
  \Sigma_{cr}\equiv  \frac{c^2}{4\pi G}D_{lens}^{-1}
\end{equation}  
represents the critical value of the two-dimensional mass density in
order to have strong lensing effects.  The quantity $D_{lens}$ is the
so-called \textit{lensing distance}, defined as
\begin{equation}
  D_{lens}\equiv\dfrac{D_{LS}D_L}{D_S}\,,
\end{equation}
where $D_S$, $D_L$ and $D_{LS}$  are the angular diameter distances of
the source,  of the  lens and between  source and  lens, respectively.
Although for elliptical  lenses we have to add the  effect of shear to
the  condition (\ref{kappa}),  we can  infer, to  first approximation,
what  are  the  system  configurations which  are  more  efficient  in
producing  strong lensing  features  by  investigating how  $D_{lens}$
changes in the different cosmological models, once the lens properties
and the  source redshifts are kept  fixed.  We remind the  reader that
$D_{lens}$ contains the full dependence  on the geometry of the system
and does not depend on $\sigma_8$, but only on $\Omega_m$.  We fix the
source position at redshift $z_s=2$ and we study $D_{lens}(z_l)$, that
is we keep fixed the length of the lensing system and we move the lens
from the observer towards the source  plane.  The results are shown in
Fig.~\ref{lensDist}: we  see that increasing the  value of $\Omega_m$,
the  strong  lensing efficiency  reaches  its  maximum at  lower  lens
redshifts.   In particular  the peak  around which  the production  of
gravitational   arcs   is  expected   to   be   boosted  shifts   from
$z\approx  0.6$ to  $z\approx 0.4$  when  the value  of $\Omega_m$  is
increased from 0.1 to 0.5.

The effect of the anticipation of  structure formation due to a higher
value of $\sigma_8$  \citep{giocoli07a,giocoli12b} has consequences on
several  halo structural  properties that  may influence  the size  of
$\sigma_{l/w}$.   Considering the  concentration  parameter, at  fixed
$\sigma_8$,  large $\Omega_m$  values  lead  to larger  concentrations
because the structures  form and grow in denser  environments.  At the
same time, keeping fixed the  value of $\Omega_m$, in cosmologies with
high $\sigma_8$ the concentration increases because of both the higher
contrast  between primordial  perturbations  and  background, and  the
anticipated formation time \citep{neto2007,giocoli12b,maccio08}.

Haloes  triaxiality  is also  an  important  feature that  depends  on
cosmological     parameters    \citep{despali2014,bonamigo15}.      In
particular,  the level  of sphericity  of  a halo,  which is  directly
related to the  ratio between its minor and major  semi-axes $a/c$, is
an  increasing function  of $\sigma_8$  and a  decreasing function  of
$\Omega_m$.  As an example, if we consider haloes with a mass equal to
$7.5\times10^{14}$ $h^{-1}\mbox{M}_{\odot}$ at  redshift $z=0.54$ in a
cosmological model  with $\Omega_m=0.3$, the median  ratio among $128$
realizations varies from $a/c=0.353_{-0.056}^{+0.049}$ in a model with
$\sigma_8=0.6$,  to  $a/c=0.417_{-0.066}^{+0.057}$  in  a  model  with
$\sigma_8=1.0$.   The  quoted  uncertaintes correspond  to  1$\sigma$
errors.   On the  other  hand,  if we  fix  $\sigma_8=0.8$, the  ratio
changes   from   $a/c=0.419_{-0.066}^{+0.058}$   in   a   model   with
$\Omega_m=0.1$,  to  $a/c=0.388_{-0.061}^{+0.053}$  in  a  model  with
$\Omega_m=0.5$.

\begin{figure}
 \centering
 \includegraphics[scale=0.4]{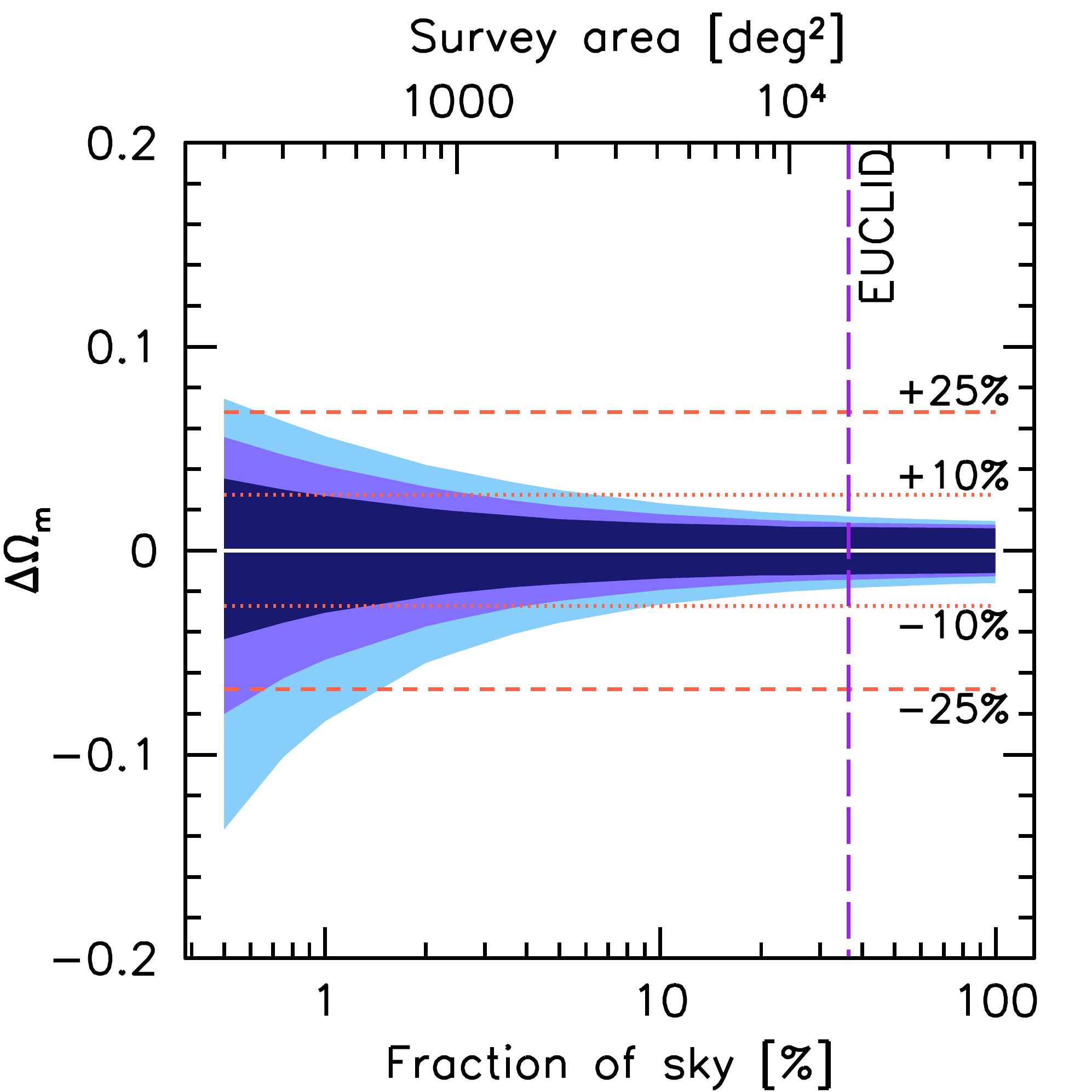}
 \caption{The amplitude of the $3\sigma$ error bar on $\Omega_m$ as a
   function of the survey area. The vertical dashed line shows the
   size of the future Euclid wide survey. The red horizontal dotted
   and dashed lines represent a variation of $\Omega_m$ corresponding
   to $\pm10\%$ and $\pm25\%$, respectively.}
 \label{biasplot}
\end{figure}

\subsection{Effects of completeness and cluster selection function}

In the following subsections we will discuss how our results change
when we take into account the lack of completeness and when we
introduce a realistic photometric galaxy cluster selection function.

\subsubsection{The effect of sample completeness and survey area}
\label{bias}
Let us consider here the case in  which a fraction of arcs are missed,
independently  of the  properties of  the lens  configurations ($l/w$,
$z_l$, $z_s$).  This may happen because some arcs may escape detection
for some  particular configurations  of the light  distribution within
the  cluster,  or  when  the  separation  between  cluster  and
foreground galaxies  is made  difficult by the  lack of  precise color
information.
The  total arc
counts may  also diminish because  we are  performing our search  in a
reduced effective area, smaller than the one of the running survey. In
this situation losing $10\%$ of the  counts is equivalent to observe a
portion of  sky $10\%$ smaller  than the original survey.  The obvious
consequence of a reduction of the  number of observed arcs is that the
Poissonian uncertainty grows and can start to dominate with respect to
the cosmic variance,  when accounting for the total  error budget.  To
quantify this effect, in Fig.~\ref{biasplot} we show, as a function of
the fraction  of the sky covered  by the arc search,  the variation of
the $3\sigma$ error bar on the parameter $\Omega_m$, when the value of
$\sigma_8$   is    a   priori    fixed   to   its    reference   value
($\sigma_8=0.809$), as  it may  happen if independently  measured from
other cosmological probes.  Dark, medium  and light blue regions refer
to the  cases of  arcs with $l/w\ge5$,  $7.5$ and  $10$, respectively,
while the horizontal dotted (dashed)  lines indicate an accuracy of 10
(25) per  cent on $\Omega_m$.  From  the figure it is  clear that arcs
with $l/w\ge5$, being more numerous,  give stronger contraints and are
less affected by possible  incompleteness problems.  However, there is
a difficulty when dealing with them  because they can look like simple
edge-on galaxies.  For  this reason the loss  and misidentification of
arcs are  expected to  depend on $l/w$,  being stronger  for low-$l/w$
ratios.   From  this point  of  view,  Fig.  \ref{biasplot}  is  quite
encouraging: if the survey area is sufficiently wide (larger than 10\%
of the whole sky), or equivalently if the arc finders are sufficiently
efficient, the error budget is  dominated by cosmic variance and there
is  not a  significant difference  in the  constraining power  between
using arcs  with $l/w\ge 5$ or  with $l/w\ge 10$.  We  remind that the
SDSS \citep{sdss2000} has  an area of about 10,000  deg$^2$, while the
Euclid   wide   survey   is   expected   to   cover   15,000   deg$^2$
\citep{EUCRB}.

\begin{figure}
 \centering
 \includegraphics[scale=0.4]{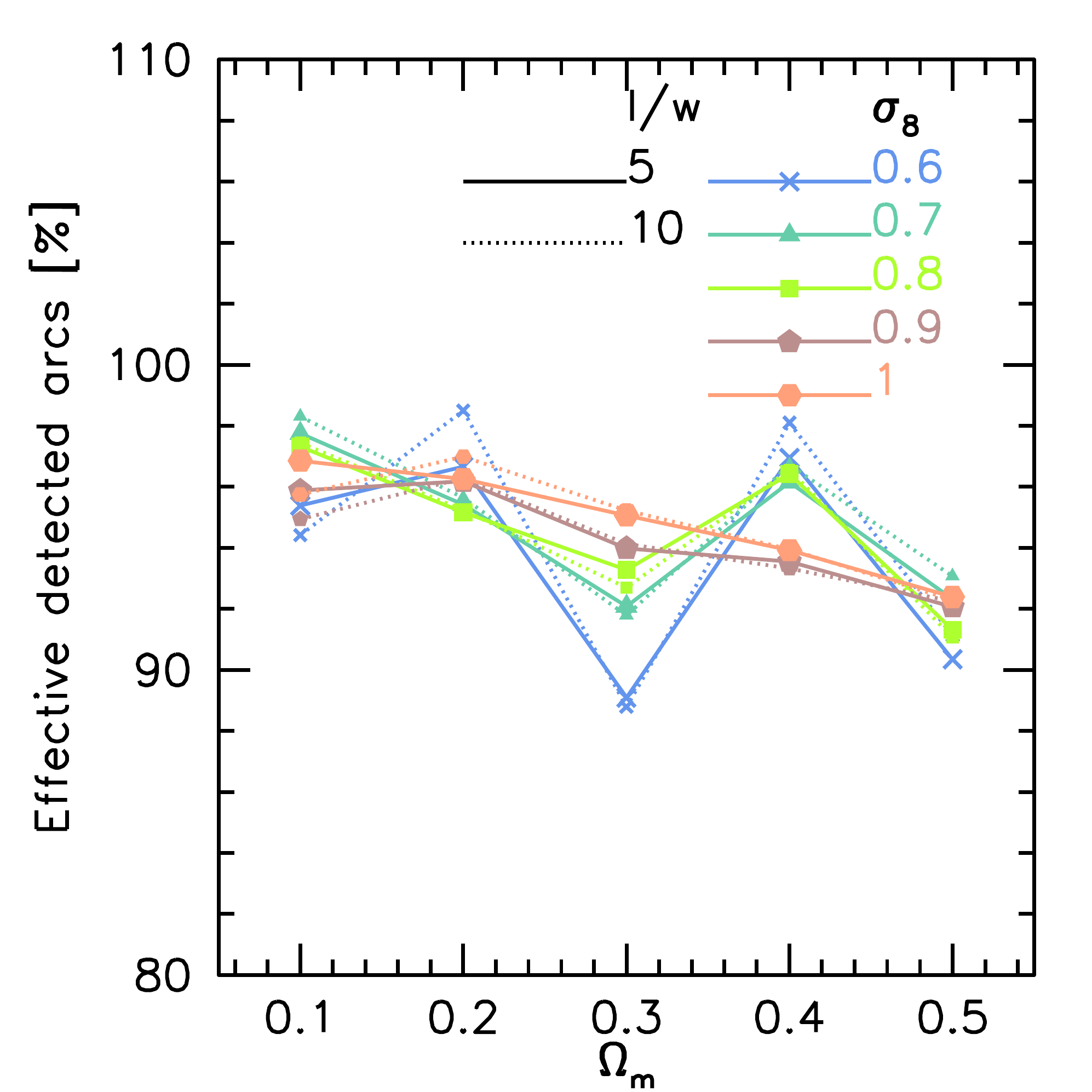}
 \caption{Percentage of arcs effectively detected by considering only
   lenses having a mass larger than the Euclid cluster photometric
   selection function. Different color refer to different values of
   $\sigma_8$, as labeled; solid and dotted lines are for arcs with
   $l/w\ge5$ and 10.}
 \label{contBIV}
\end{figure}

\begin{figure}
 \centering
 \includegraphics[scale=0.50,angle=270]{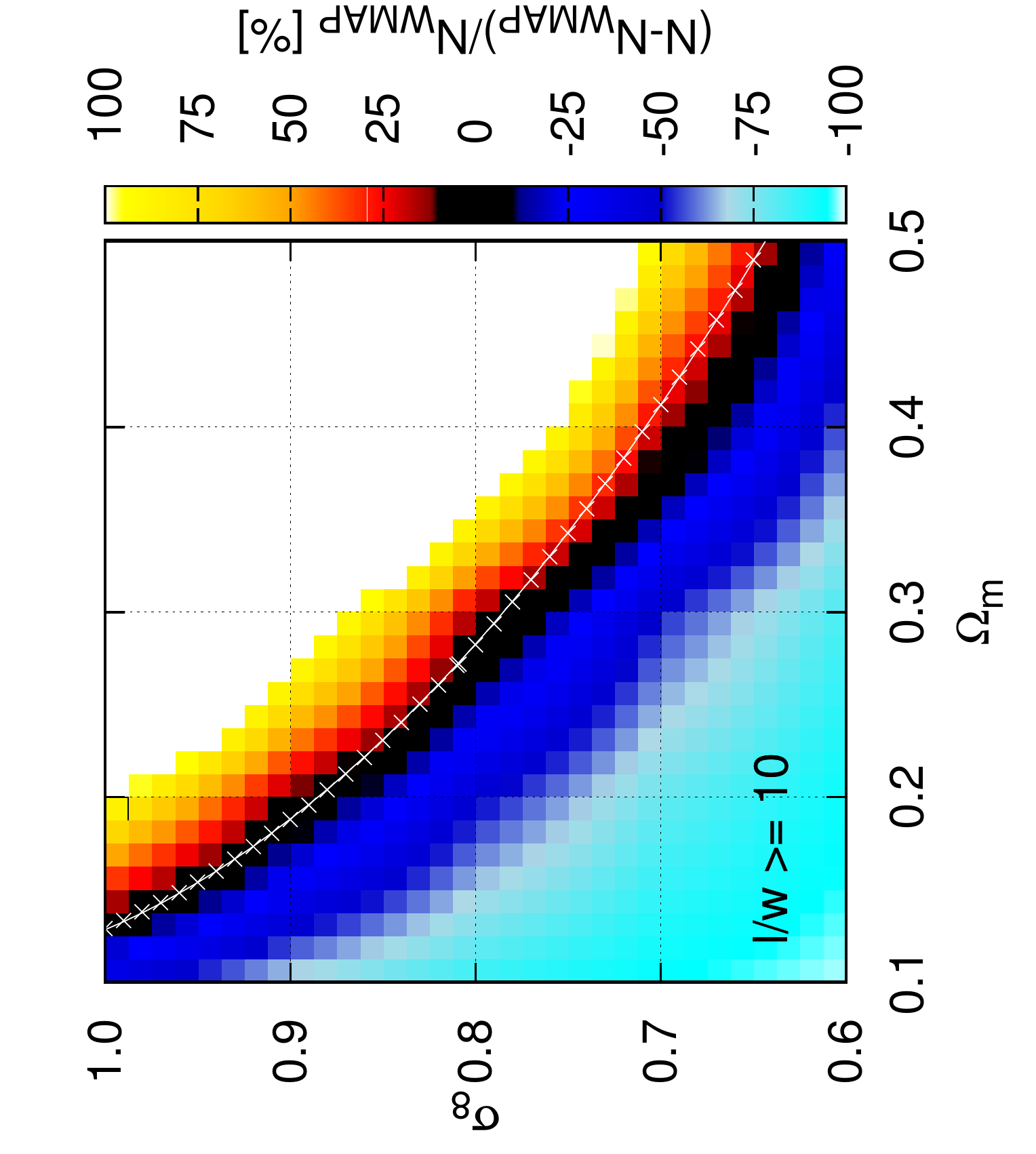}
 \caption{ As the  upper plot of Fig.~\ref{constraints},  but for arcs
   produced by  lenses having  a mass larger  than the  Euclid cluster
   photometric selection function \citep{sartoris15}. The crossed line
   represents the degeneracy curve obtained when no selection function
   is applied.}
 \label{contBIV2}
\end{figure}

\subsubsection{The effect of the cluster selection function}
\label{selfct}

Due to the high computational cost of the algorithms for arc
detection, a possible strategy in future wide surveys is to run these codes only on
small-size frames where galaxy clusters have been previously
identified. Obviously, this originates a reduction of the effective
number of arcs, which is strongly dependent on the specific cluster
selection function of the survey.

As a  worked example,  here we  consider again  the future  ESA Euclid
mission.  Given the amount and quality of its data covering an area of
15,000 deg$^2$,  there will be  at least  three main ways  to identify
galaxy clusters:  (i) from  photometric data, (ii)  from spectroscopic
data,   and   (iii)   from   cosmic   shear   maps.    As   shown   in
\citet{sartoris15} , the one based on photometric data \citep[see, for
  example,][and  references therein]{bellagamba11}  is expected  to be
largely the  most efficient one.   In this  case, the minimum  mass of
galaxy clusters  having a number  of members  larger than 3  times the
r.m.s.  of the field galaxy counts  is expected to be between $5\times
10^{13}\,M_{\odot}/h$  and   $8\times  10^{13}\,M_{\odot}/h$   in  the
redshift range  here considered  \citep{sartoris15}.  Compared  to the
minimum mass needed  to produce critical lines for  sources located at
redshift  $z_s=2$   (see  Fig.~\ref{contBIV0}),  the   Euclid  cluster
selection is  then slighly  higher on a  limited redshift  range only,
namely  between $z=0.2$  and $z=0.5$.   This means  that limiting  the
search  for arcs  to frames  where galaxy  clusters have  been already
identified  is  expected to  not  reduce  dramatically the  number  of
detected arcs. This is confirmed  in Fig.~\ref{contBIV}, where we show
the fraction of  arcs that can be effectively  detected following this
strategy. Same colors  indicate same values of  $\sigma_8$, as labeled
in the figure, while solid and dotted lines refer to arcs with $l/w\ge
5$  and   $10$,  respectively.   For  the   cosmological  models  here
considered, the  reduction varies  between 2  and 10  per cent  and is
almost  independent of  $l/w$.   For the  reference  WMAP7 model,  the
percentage of effectively detected arcs remains about 95 per cent.

In Fig.  \ref{contBIV2} we show the relative differences in the arc
counts between each cosmological model and the reference WMAP7 cosmology,
considering only arcs produced by galaxy clusters above the Euclid
photometric selection function.  In the figure, the color scale is
identical to that adopted in the upper panel of Fig.~\ref{constraints}.  The white
crosses represent the degeneration curve we found considering the
total number of arcs, i.e.  without applying the cluster selection
function.  Although similar, the curve changes in a non negligible
way, especially considering extreme values of the parameters.  This
underlines the importance of taking into account every kind of
selection function when combining theory and observations in arc
statistic studies, avoiding possible systematics.

\begin{figure}
  \centering
  \includegraphics[scale=0.4]{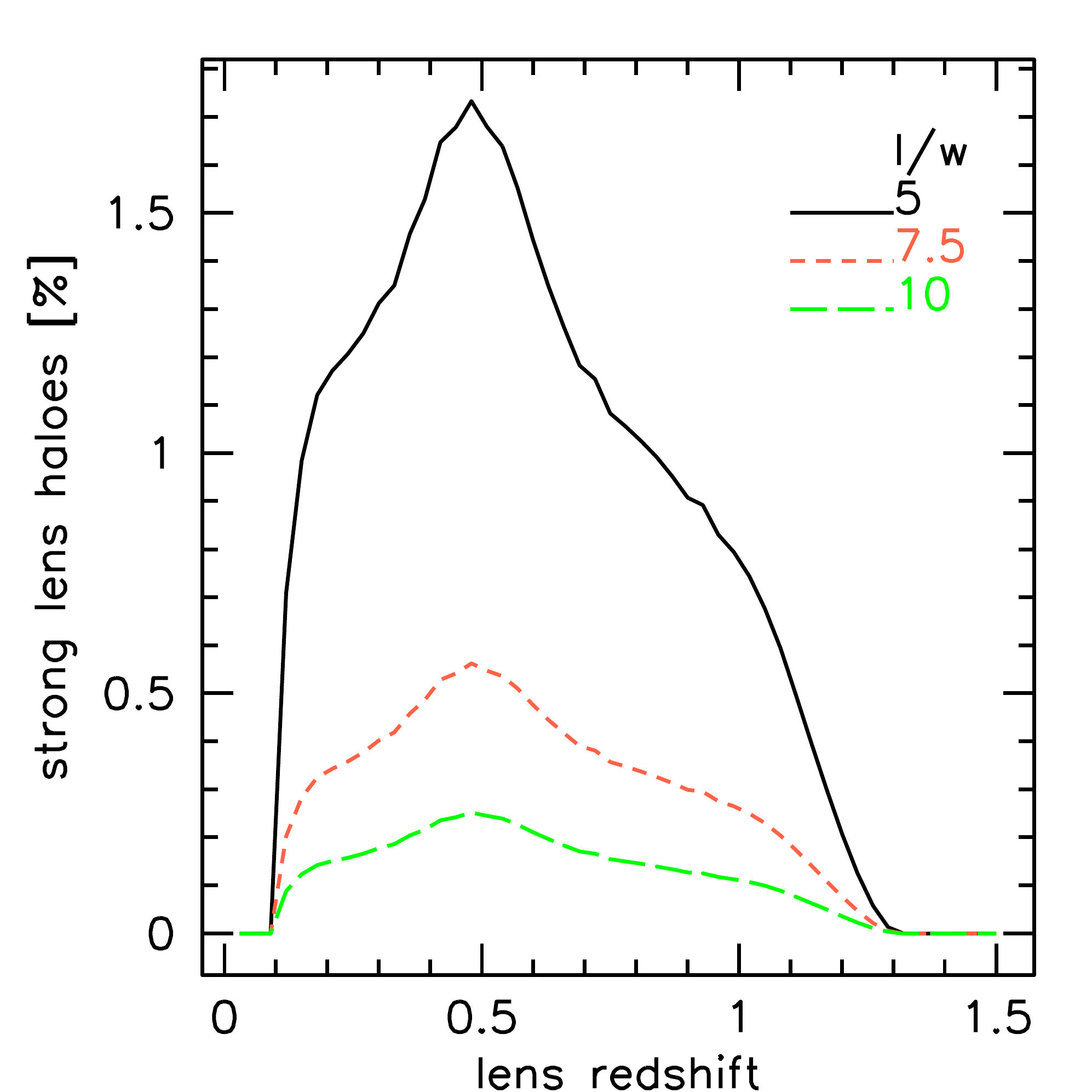}
  \caption{ Fraction of galaxy clusters having a mass larger than the
    Euclid cluster photometric selection function producing at least
    one giant arc. Results are shown for the reference WMAP7 model.
    Different line styles refer to different length-to-width ratios,
    as labeled. }
  \label{contBIV3}
\end{figure}

The  presence of  strong lensing  features like  arcs can  represent a
complementary   way   to   confirm    the   presence   of   a   galaxy
cluster. Moreover  arcs can be  used to  improve the estimates  of the
mass of galaxy clusters, a fundamental ingredient to fully exploit the
evolution of their abundance as cosmological probe. For this reason it
is important  to compute what is  the fraction of the  galaxy clusters
identified in the Euclid photometric survey, which are able to produce
at least one  giant arc. The result for the  reference WMAP7 cosmology
as a  function of  redshift is shown  in Fig.~\ref{contBIV3}  for arcs
with  $l/w\ge 5$,  7.5  and  10 (black  solid,  red  dashed and  green
long-dashed lines,  respectively).  Typical  mean values are  around 1
per cent, 0.33 per cent and 0.15  per cent for $l/w\ge 5$, 7.5 and 10,
respectively.  From the figure we notice that the strong lens fraction
peaks around $z=0.5$: this behaviour  is a combined effect between the
well of  the photometric  selection function around  redshift $z=0.75$
and  the peak  -- around  the  same redshifts  -- of  the strong  lens
counts. Interestingly, for redshift  $z\ge1.3$ the percentage tends to
vanish.   Finally we  notice that  the  fact that  the strong  lensing
selection function  can be  smaller than  the photometric  cluster one
would allow  in principle to add  extra objects to the  Euclid cluster
sample by looking for strong lensing features only. However this would
require to run  the algorithms for arc detection  blindly in different
areas of the  survey.  Considering the reference WMAP7  model and arcs
with $l/w\ge 5$, the gain  would correspond to approximately 300 extra
objects    only,    all    having   a    relatively    low    redshift
($0.2\le z \le 0.5$).

Therefore,  we   can  conclude   that  arc  statistics   represents  a
complementary tool to identify galaxy  clusters or eventually to prove
their presence.  In particular, arcs with  a small $l/w$ ratio are the
best tracers,  since they are  more numerous,  but, at the  same time,
they are  the more difficult  to identify because of  their similarity
with  non-lensed galaxies.   Finally, our  results underline  that the
codes for arc identification can be  run on single frames where galaxy
clusters  have  been already  detected  with  no consequences  on  the
cosmological predictive power of arc statistics.

\section{A test-bed for the method: the CLASH survey}
\label{testbed}
While this paper focuses on the sensitivity of arc statistics to cosmological parameters like $\Omega_m$ and $\sigma_8$, it is worth mentioning that another paper has been recently submitted by our collaborators \cite{xu15} to compare theoretical predictions of arc abundances in a $\Lambda$CDM cosmological model and observations. More precisely, in this other work MOKA has been used to build up halos reproducing the properties  of the X-ray selected galaxy clusters belonging to the CLASH sample \citep{postman12}. Numerical hydro-dynamical simulations tailored to reproduce the CLASH selection function \citep{meneghetti14} are also used to derive theoretical predictions. Thus, the work of \cite{xu15} provides the best opportunity for validating our methodology against more complex models of the cluster mass distribution and against observed clusters with a known selection function. 

The results of this study show that there is an excellent agreement between expectations based on MOKA halos and numerical simulations and the arc counts in the CLASH clusters. More specifically, the lensing efficiency measured in the CLASH sample is $4 \pm 1$ arcs (with $l>6"$ and $l/w>7$) per cluster. MOKA simulations return exactly the same number ($4 \pm 1$), while numerical simulations give $3 \pm 1$ arcs per cluster. Therefore, according to \cite{xu15}, in terms of efficiency to produce long and thin arcs, observations and simulations based on MOKA and numerical hydro-dynamical techniques come into full agreement. It is particularly significant  that the methodology we have developed for modeling cluster lenses for arc-statistics calculations is fully capturing the complexity of numerically simulated halos, as evinced from the fact that the cross sections for giant arcs of MOKA generated halos are well matching those of the halos described in \cite{meneghetti14}.

\section{Conclusions}
\label{conclusions}
In this work we have investigated how the number of gravitational arcs
depends on  cosmology, focusing  our attention  on the  (total) matter
density parameter  $\Omega_m$ and  on the initial  normalisation power
spectrum parameter $\sigma_8$.  In more detail we  have considered the
ranges  $\Omega_m=[0.1-0.5]$   and  $\sigma_8=[0.6-1.0]$.    Our  main
results can be summarised as follows.

\begin{itemize}
\item We confirm that arc statistics is very sensitive to the couple
  of parameters $\Omega_m-\sigma_8$. In particular we find that the
  expected number of arcs is an increasing function of both
  parameters: this is mostly due to the fact that increasing these
  parameters boosts the number of lenses.
\item The efficiency in producing arcs in cosmologies with high values
  of $\sigma_8$ is larger, since it has an effect also on the
  structure formation time, that in turn affects some lens structural
  properties (mainly concentration and triaxiality) relevant for
  strong lensing.
\item A strong degeneracy exists between the two considered
  cosmological parameters for the number of arcs $N_{arcs}$; for the
  reference WMAP7 model this is
  expressed by the relation~(\ref{totfit}), that is similar, but not
  equal, in shape to the degeneracy derived from galaxy cluster counts (see
  eq.~\ref{halofit}).  The differences between the two arise from the
  non negligible contribution to $\sigma_{l/w}$ given by the lens
  structural properties -- triaxiality, asymmetries, concentration,
  substructures and the BCG -- and the lensing distance relation.
\item  Arcs with  small $l/w$  ratio  are more  suitable to  constrain
  cosmological parameters, since they are  more numerous. On the other
  hand,  they could  be  more difficultly  identified  in the  surveys
  because of their similarity with  non-lensed galaxies.  We find that
  if the  survey area is  sufficiently larger  (more than 10\%  of the
  full sky) the error budget is  dominated by cosmic variance, and the
  constraining power of  arc counts becomes almost  independent of the
  value of $l/w$.  In particular a survey covering 15,000 deg$^2$ will
  be  able  to  distinguish  at  more than  $5\sigma$  level  the  two
  cosmological models supported by WMAP7 and Planck CMB data.
\item Considering future wide surveys, like the ESA Euclid mission, we
  find that  searching for arcs  only in frames where  galaxy clusters
  have been previously detected will produce  a loss of 2-10\% of arcs
  only (depending on the cosmological  model) and a consequent limited
  degradation of the  constraining power of arc  counts. This suggests
  that it will  be not necessary to run  the computationally expensive
  algorithms for arc detection on whole wide surveys.
\end{itemize}

In this paper we have discussed the potentiality and the capability of
the  giant arc  statistic  to  constrain the  matter  density and  the
initial power spectrum  normalisation parameter in light  of the large
data-sets that will become available from future wide-field surveys.

\section*{AKNOWLEDGEMENTS}
We  thank Barbara  Sartoris,  Cosimo Fedeli  and  Peter Schneider  for
useful  discussions on  the Euclid  cluster selection  function.  CG's
research  is part  of the  project GLENCO,  funded under  the European
Seventh Framework  Programme, Ideas,  Grant Agreement n.   259349. CG thanks
CNES for finantial support. LM acknowledge   financial   contributions   from   contracts   ASI/INAF/I/023/12/0, by  the PRIN  MIUR 2010-2011 ``The  dark Universe  and the cosmic evolution of  baryons: from current surveys to  Euclid'' and by
the PRIN INAF 2012. We are greatful to the referee Prasenjit Saha for his useful comments.
\bibliographystyle{mn2e}
\bibliography{Boldrin_cosmoarcstat}

 \label{lastpage}
\end{document}